\documentclass[aps,letter,prb,twocolumn]{revtex4-1}	
	\usepackage{amssymb}
	\usepackage{amsmath}
	\usepackage{amsthm} 
	\usepackage{bm}
	\usepackage{enumerate}
	\usepackage{graphicx}
	\usepackage[caption=false]{subfig}
	\usepackage[version=3]{mhchem}
	\usepackage{dcolumn} 
	\usepackage[colorlinks,citecolor=blue,linkcolor=blue,urlcolor=blue]{hyperref}
	
	\newcommand{\avg}[1]{\left< #1 \right>} 
	\newcommand{\trace}{\mbox{Tr}} 
	
\begin{document}
	\title{Noncollinear magnetic ordering in a frustrated magnet: Metallic regime and the role of frustration}
	\author{Munir~Shahzad and Pinaki~Sengupta}
	\affiliation{School of Physical and Mathematical Sciences, Nanyang Technological University, 21 Nanyang Link, Singapore 637371}
	\date{\today}
	\begin{abstract}
    We explore the magnetic phases in a Kondo lattice model on the geometrically frustrated Shastry-Sutherland lattice at metallic electron densities, searching for noncollinear and noncoplanar spin textures. Motivated by experimental observations in many rare-earth-based frustrated metallic magnets, we treat the local moments as classical spins and set the coupling between the itinerant electrons and local moments as the largest energy scale in the problem. Our results show that a noncollinear flux state is stabilized over an extended range of Hamiltonian parameters. These spin states can be quenched efficiently by external fields like temperature and magnetic field as well as by varying the degree of frustration in the electronic itinerancy and exchange coupling between local moments. Interestingly, unlike insulating electron densities that we discussed in paper I of this sequence,  a Dzyaloshinskii-Moriya interaction between the local moments is not essential for the emergence of their noncollinear ordering.
	\end{abstract}
	\maketitle
	
	\section{Introduction}\label{sec:intro}
	
The interplay between charge and spin degrees of freedom results in various novel and exotic phases in strongly correlated electron systems. A paradigmatic model to describe this interplay is the Kondo lattice model (KLM) or double-exchange model (DEM), in which localized moments are coupled to itinerant electrons~\cite{zener-1951,anderson-1955,furukawa-1995,yunoki-1998,kubo-1972}. The mobile electrons in these systems act as mediators to establish the effective correlation between localized spins, giving rise to magnetic behavior. On the other hand, the scattering of these electrons from localized moments may affect the electronic and transport properties of these systems. The study becomes more fascinating when we have geometrical frustration as an extra degree of freedom. In such metallic frustrated magnets, the localized spins are arranged on some geometrically frustrated lattice. These charge-spin coupled systems on frustrated lattices often exhibit unconventional noncollinear spin textures that are not observed in their nonfrustrated counterparts and drive novel and exotic topological properties ~\cite{grohol-2005,balicas-2011,wen-1989,wen-1991,laughlin-1990}. For noncoplanar spin configurations, a measure of the noncoplanarity is a nonzero value of the scalar spin chirality defined on a triangle $\chi_\bigtriangleup=\mathbf{S}_i\cdot\mathbf{S}_j\times\mathbf{S}_k$. A nonzero value of chirality on frustrated lattices breaks both time-reversal $\mathcal{T}$ and parity $\mathcal{P}$ symmetries and drives some peculiar transport phenomena such as geometrical or topological Hall effect
	(THE)~\cite{udagawa-2013,boldrin-2012,taguchi-2001,nagaosa-2001,machida-2010,machida-2007}. These chiral spin configurations act as a source of fictitious magnetic field; when an itinerant electron moves over them in a closed path, it picks up an extra Berry phase, which results in THE~\cite{jinwu-1999,berry-1984,daniel-1992,kenya-2000}. 
THE has been studied for Ferromagnetic Kondo lattice model on geometrically frustrated lattices such as triangular~\cite{yasu-2010,kipton-2013,rahmani-2013,takatsu-2010}, kagome~\cite{kenya-2000,kipton-2014,chern-2014}, checkerboard~\cite{venderbos-2012}, pyrochlore~\cite{chern-2010,nakatsuji-2006}, and fcc~\cite{shindou-2001} lattices. Chiral spin textures are essential to realizing these novel transport phenomena. In addition to the chiral spin textures, some other examples of unconventional magnetic order include the flux state, in which the spins are coplanar and arranged in a cyclic pattern on a square plaquette, and the
``all-in, all-out" state on the kagome lattice, which also supports novel transport phenomena~\cite{chen-2014,kubler-2014,nayake-2016}. 
		
The unconventional spin textures emerging dynamically from the interplay between competing microscopic interactions on a frustrated lattice can be controlled with the help of external magnetic field, temperature, and pressure, which make them a potential candidate for application in spintronics. In this regard, it is very desirable to study the nature of these magnetic ordered states and associated phase transitions in order to stimulate experimental work. In an accompanying study~\cite{shahzad-2017}, we have investigated the role of Dzyaloshinskii-Moriya (DM) interactions on stabilizing chiral spin textures at half filling, where there exists a finite gap in the electronic spectrum. We found that for the insulating state, DM interactions are essential in establishing the noncoplanar ordering of the local moments. In this work, we focus on electronic filling factors $n_e={1\over 4}$ and ${3\over 4}$ (where $n_e=\frac{1}{2N}\sum_{i\sigma} \avg{c_{i\sigma}^\dagger c_{i\sigma}}$), for which the electronic spectrum is gapless; that is, the ground state is metallic. We demonstrate that at these parameter regimes, unconventional noncollinear spin textures emerge even in the absence of DM interactions. Our results reveal that while noncoplanar ground states are not realized in the absence of DM interactions, the noncollinear flux state is stabilized for a wide range of parameters involved in the Hamiltonian. We discuss in detail the nature of the magnetic ground states and associated phase transition as a function of thermal fluctuations, frustration and external magnetic field for both number densities of electrons.

\begin{figure}[htb]
	\centering	  
	\subfloat[]{\label{fig:SSL}{\includegraphics[scale=0.5,trim=0cm 0cm 0cm 0cm,clip]{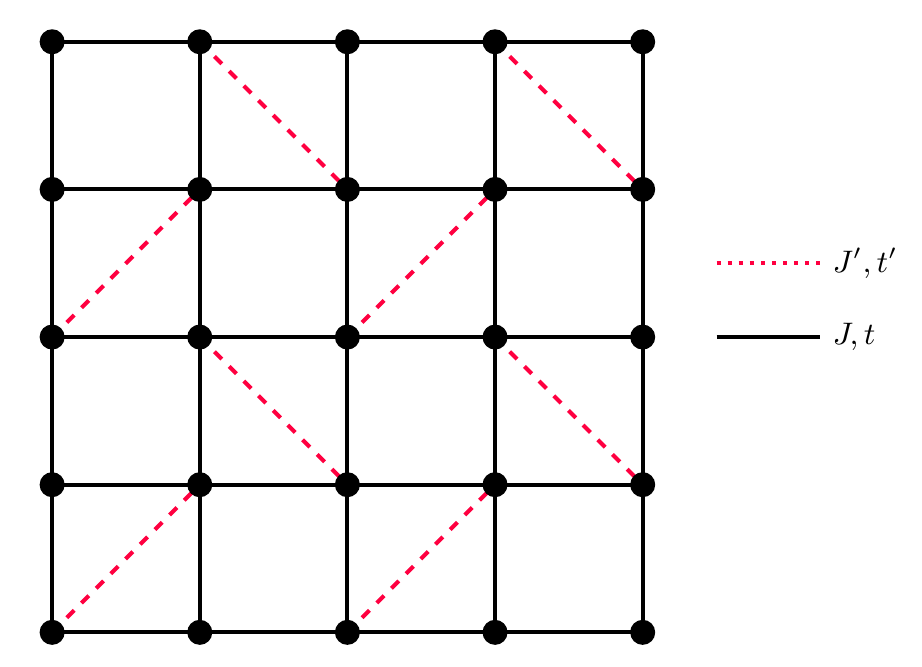}}}
	\subfloat[]{\label{fig:SSL-BZ}{\includegraphics[scale=0.5,trim=0cm 0cm 0cm 0cm,clip]{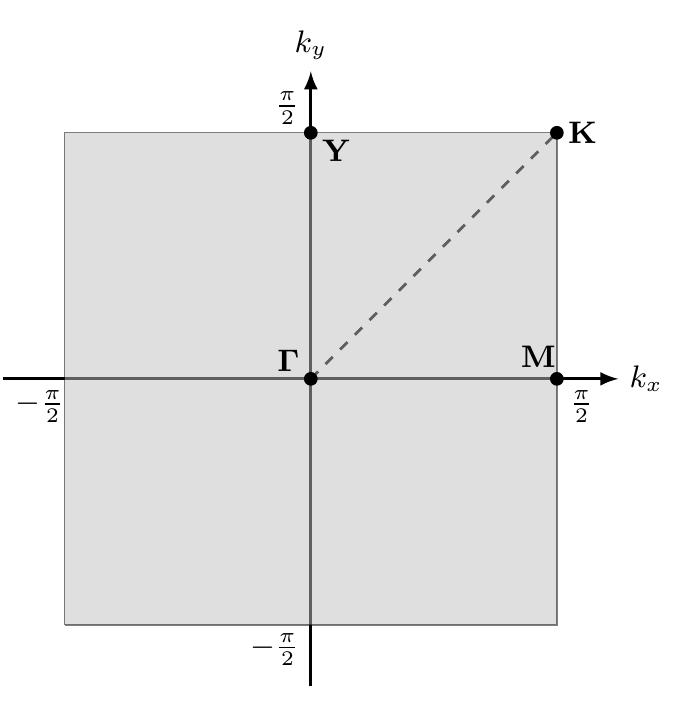}}}
	\caption{(Color online) (a) The geometry of SSL, where there are diagonal bonds on alternate plaquettes in addition to axial bonds along the $x$ and $y$ axes. (b) First BZ of the SSL lattice with high-symmetry points indicated.}
	\label{fig:ssl06}
\end{figure}

    \section{Model}\label{sec:model}

    We consider the Shastry-Sutherland Kondo lattice model (SS-KLM) in the presence of a longitudinal magnetic field. The geometry of the Shastry-Sutherland lattice (SSL) along with the first Brillouin zone (BZ) is depicted in Fig.~\ref{fig:ssl06}. The Hamiltonian describing the system under investigation is
	
	\begin{multline}
	\label{equ:ham}
	\mathcal{\hat{H}}= \underbrace{-\sum_{\avg{i,j},\sigma}t_{ij}(c_{i\sigma}^\dagger c_{j\sigma}+\mbox{H.c.})-J_K \sum_i \mathbf{S}_i\cdot \mathbf{s}_i}_{\mathcal{\hat{H}}_e}\\ \underbrace{ +\sum_{\avg{i,j}}J_{ij}\mathbf{S}_i \cdot \mathbf{S}_j -h^z\sum_i S_i^z }_{\mathcal{\hat{H}}_c}.
	\end{multline} 
	The first two terms make up the electronic part of the Hamiltonian $\mathcal{\hat{H}}_e$, where we have a tight-binding term for the itinerant electron and an on-site Kondo-like interaction between the spin of the itinerant electron $\mathbf{s}_i=c_{i,\alpha}^\dagger\bm\sigma_{\alpha\beta}c_{i,\beta}$ (where $\bm\sigma_{\alpha \beta}$ are the vector elements of the usual Pauli matrices) and localized spin $\mathbf{S}_i$. Here $\avg{i,j}$ denotes the bonds on the SSL, where nearest neighbors (NN) are axial bonds and next-nearest neighbors (NNN) are diagonal bonds on the alternate plaquettes and $t_{ij}$ describes the hopping matrix of itinerant electrons on these SSL bonds. We treat the localized spins $\mathbf{S}_i$ as classical vectors of unit length, so the sign of $J_K$ becomes irrelevant in the current model since eigenstates that correspond to different sign of $J_K$ are related by a global gauge transformation ~\cite{pekker-2005,martin-2008}. The third term in Eq.~\eqref{equ:ham} represents antiferromagnetic Heisenberg interaction between the localized spins. The last term in the Hamiltonian~\eqref{equ:ham} is the Zeeman term for the localized spins due to an external (longitudinal) magnetic field. We consider the strong $J_K$ coupling limit; in that case, the spin direction of the itinerant electron is completely determined by the localized moment. So we consider the effect of an external magnetic field on only the localized moments. From here onwards, we represent the interactions on the axial bonds as unprimed parameters, while primed parameters are used for diagonals bonds. We choose $t=1$ as the energy unit.   
	
	\section{Method and observables}\label{sec:methods}
	
The above model can be explored for thermodynamic properties using an unbiased Monte Carlo (MC) method~\cite{motome-1999,furukawa-2004,ishizuka-2012,yasu-2010,alvarez-2006,ishizuka-2013,ishizuka-2015}. We have already discussed this approach in the first paper of this series; here we 
present a brief outline for completeness. The slow dynamics of localized moments can be decoupled from the fast dynamics of itinerant electrons by treating them as static classical fields on each site. Replacing the itinerant electron spin in terms of raising and lowering operators of itinerant electrons, the electronic part of the Hamiltonian $\mathcal{\hat{H}}_e$ becomes quadratic in fermionic operators. The one-electron basis can be used to express this Hamiltonian as a $2\mathit{N}\times 2\mathit{N}$ matrix for a particular arrangement of classical localized spins. The full partition function in the grand-canonical ensemble can be represented in terms of two traces: $\trace_{c}$ over the classical localized moments $\{x_r\}$ and $\trace_f$ over fermionic degrees of freedom. The eigenvalues of the Hamiltonian matrix $\mathcal{\hat{H}}_e$ for a fixed configuration of the localized spins can be used to calculate the trace $\trace_f$. The partition function then can be written as,
    \begin{equation}
	\label{equ:part func}
	\mathit{Z} =\trace_c \exp[-\mathit{S_{eff}}(\{x_r\})-\beta (\mathcal{\hat{H}}_c)],
	\end{equation}
	where $\mathit{S_{eff}}(\{x_r\})=\sum_\nu\mathit{F}(y)$ is the effective action and $\mathit{F}(y)=-\ln[1+\exp\{-\beta(y-\mu)\}]$. The number density of itinerant electrons is adjusted through the chemical potential $\mu$, and $\beta=1/k_BT$ represents the inverse temperature. To calculate $\trace_c$ a classical MC method is used to sample the spin configuration space.
									
	The eigenvalues and eigenfunctions of $\mathcal{\hat{H}}_e(\{x_r\})$ are used to calculate the thermodynamic quantities related to itinerant electrons, while the quantities associated with the localized spins are calculated with the thermal averages of spin configurations. We select a random configuration of localized spin $\{x_r\}$ and calculate the Boltzmann action $\mathit{S_{eff}}(\{x_r\})$ for this configuration. Next, the random updates performed for this spin configuration are accepted or rejected based on the Metropolis algorithm. We use the static spin structure factor to distinguish between different magnetic orders of the localized spins,
			
	\begin{equation}
	\label{equ:str fact}
	S(\mathbf{q})=\frac{1}{N}\sum_{i,j}\avg {\mathbf{S}_i \cdot \mathbf{S}_j} \exp [i\mathbf{q} \cdot \mathbf{r}_{ij}],
	\end{equation}
	where $ \avg{\cdot} $ represents the thermal averages over the grand-canonical ensemble and $\mathbf{r}_{ij}$ is the position vector from site $\mathit{i}$ to $\mathit{j}$. We calculate the uniform magnetization per site,
		 
    \begin{equation}
	\label{equ:magnet02}
    m=\sqrt{\avg{\left (\frac{\sum_i \mathbf{S}_i}{N}\right )^2}},
    \end{equation}
    as well as staggered magnetization per site,
    \begin{equation}
    \label{equ:staggmagnet02}
    m_{stagg}=\sqrt{\avg{\left (\frac{\sum_i (-1)^i\mathbf{S}_i}{N}\right )^2}},
    \end{equation}
    to describe the evolution under varying magnetic field and frustration in electronic itinerancy, respectively.
    
	\section{Results and Discussion}\label{sec:res-disc}
	
	 We perform simulations on lattice sizes $L=8-18$, where $L$ is 
the length of the lattice along one axis. The results for $L=8$ are obtained by diagonalizing the full Hamiltonian to calculate the Boltzmann factor as it is faster for small lattice sizes. But for lattice size $L>8$, we use the traveling cluster approximation (TCA)~\cite{kumar-2006,anamitra-2015,majumdar-2006,kumar-2005}; in this method, a cluster of $6\times6$ sites is moved sequentially over the whole lattice, and the Boltzmann factor is calculated by just diagonalizing this cluster Hamiltonian. Once the system is equilibrated, the physical observables are calculated by diagonalizing the full Hamiltonian. To approach equilibration efficiently, we use a simulated annealing method. At the beginning, we select a random configuration of the localized spins at a relatively high temperature $T=0.1$ and equilibrate the system. Next, we use the final configuration of this temperature to perform the equilibration at $T=0.08$. This process is repeated with a temperature step $\Delta T=0.02$ when $T>0.01$ and $\Delta T=0.002$ when $T<0.01$, finally calculating the thermal averages of the physical observables at $T=0.005$. Measurements are made after every $1000$ steps of an MC run of $50\:000$ steps in total after discarding $60\:000$ MC steps for thermalization. We divide the data into $50$ different bins to calculate the average values and errors from the standard deviation. Moreover, the MC results are obtained using periodic boundary conditions.

	\begin{figure}[htb]
		\centering
		\includegraphics[width=0.49\textwidth]{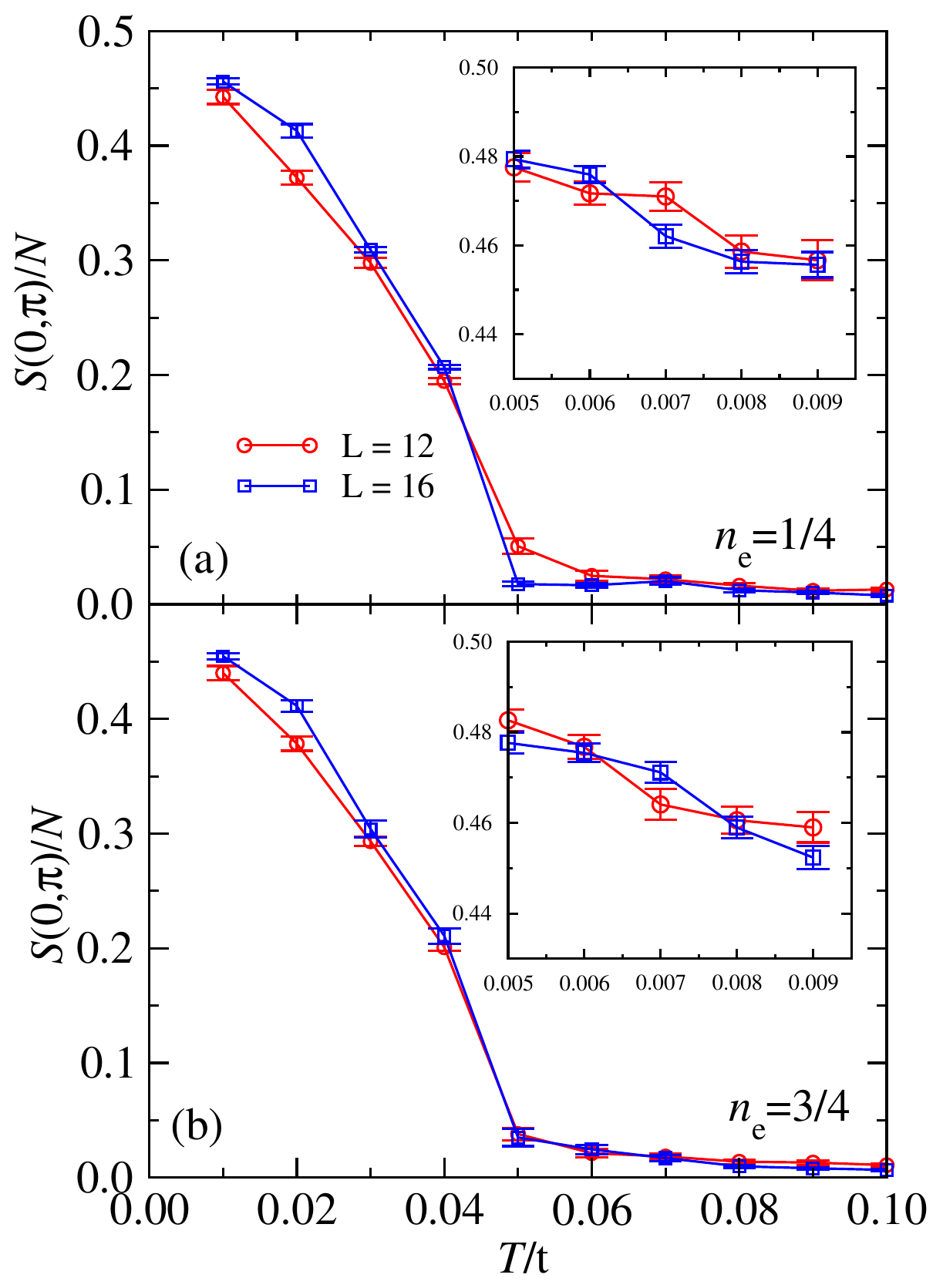}
		\caption{(Color online) The magnitude of the peak in spin structure factor at $\mathbf{q}=(0,\pi)$ plotted as a function of temperature $T$ at (a) $n_e\simeq1/4$ and (b) $n_e\simeq3/4$ for $12\times12$ and $16\times16$ lattice sizes. The parameters used are $t'/t=0.25$, $J/t=0.1$, $J'/t=0.025$, $h^z/t=0$, and $J_K/t=8.0$. Insets show the temperature dependence when $T<0.01$.}
		\label{fig:temp-dep}
	\end{figure}
	
	\begin{figure}[!htb]
		\centering
		\includegraphics[width=0.49\textwidth]{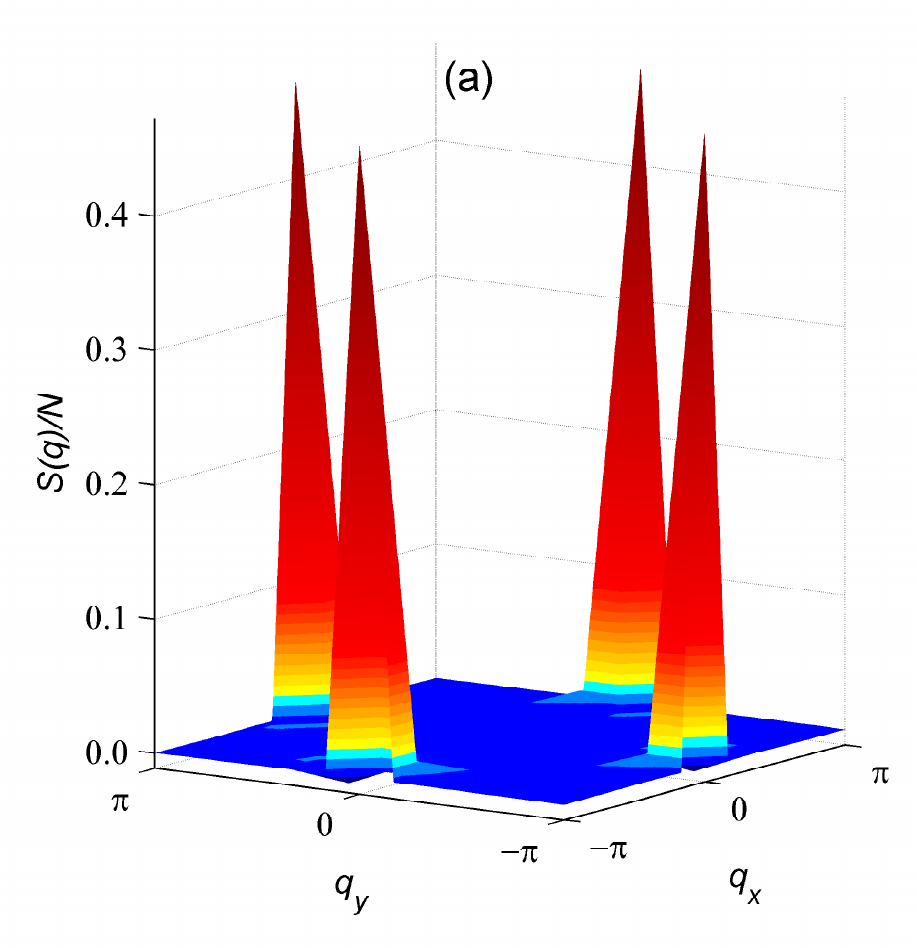}
		
		\includegraphics[width=0.49\textwidth]{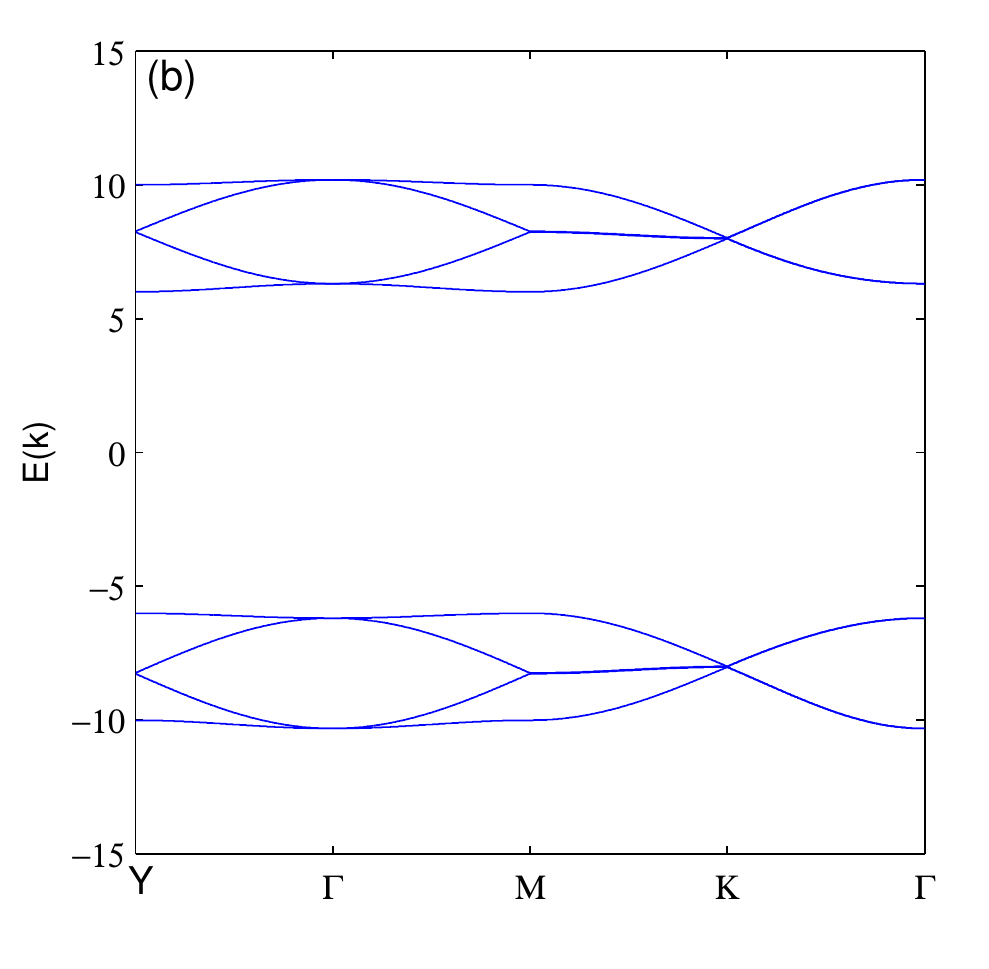}
		\caption{(Color online) (a) The momentum dependence of the static spin structure factor exhibiting two sharp and equal-magnitude peaks at $\mathbf{q}=(0,\pi)$ and $(\pi,0)$, indicating a noncollinear flux state. The results are obtained using parameters $t'/t=0.25$, $J/t=0.1$, $J'/t=0.025$, $h^z/t=0$, $J_K/t=8.0$, and $T/t=0.005$ for a $12\times12$ lattice at $n_e\simeq1/4$. (b) Electronic energy bands for a tight-binding model including double-exchange interaction with 2-$\mathbf{q}$ order stabilized at $\mathbf{q}=(0,\pi)$ and $(\pi,0)$. The values of the interaction parameters used are $t'/t=0.25$ and $J_K/t=8.0$. }
		\label{fig:sq3d-band}
	\end{figure}  
 
	\subsection{Role of temperature}
	
Throughout this work we choose the limit of strong
Kondo coupling between itinerant electrons and localized
spins $(J_K=8.0)$ following the experimental observations
in relevant metallic magnets. We start our discussion by
considering the thermal transition of the magnetic ground
state. Figure~\ref{fig:temp-dep}(a) shows the temperature dependence of the magnitude of the peak in the spin structure factor at $\mathbf{q}=(0,\pi)$ as a function of temperature at $n_e\simeq1/4$ (here we introduce the $\simeq$ sign because we are working in the grand-canonical ensemble and electron densities are controlled by the values of $\mu$ that give number densities fluctuating within the error bars close to the required one). Initially, the magnitude of
the peak at $\mathbf{q}=(0,\pi)$ remains constant and vanishingly small with decreasing temperature, but around $T\simeq0.05$ here is
a sharp increase in the magnitude of the peak indicating a
first-order magnetic phase transition. Insight into the nature
of the magnetic ordering after the phase transition can be gained from a plot of the static spin structure factor $S({\bf q})$ as a function of the momentum [see Fig.~\ref{fig:sq3d-band}(a)]. It consists of two sharp and equal magnitude peaks at 
${\bf q} = (0,\pi)$ and $(\pi,0)$. The dual peaks at $(0,\pi)$ and $(\pi,0)$ are a signature of a noncollinear flux state in similar models~\cite{yamanaka-1998,agterberg-2000}. The in-plane components of the local moments are arranged in a flux pattern. The noncollinear magnetic ordering breaks the continuous symmetry much like how a FM state breaks $\ce{O(_3)}$ even when the Hamiltonian does not contain a symmetry-breaking term. For noncollinear magnetic orderings such as the flux state the order parameter degeneracy manifold is $\ce{SO(_3)}$, i.e., a triad of unit vectors. So the flux-pattern
does not need to be in the $xy$ plane of the lattice. All the interesting results depend only on relative orientations of NN and NNN spins. Even if one considers an ordered state that is globally rotated with respect to the specific flux state, the same physics is obtained. Quenching the thermal fluctuations drives the system from a paramagnetic state with vanishingly small magnetization (due to finite-size effects) 
to a noncollinear magnetic state 
marked by a 2$\mathbf{q}$ order in the spin structure factor. In the ordered state, the local moments are oriented with their transverse 
components forming a flux pattern around the squares with no diagonal, with alternate clockwise and counterclockwise circulation around neighboring plaquettes. This flux state is stabilized due to the competition between antiferromagnetic (AFM) superexchange interaction and the double-exchange interaction (which favors the FM ordering at this filling factor) in the presence of geometrical frustration. The competition between these interactions on geometrically frustrated lattices is related to subdominant interactions such as the antiferroic biquadratic interaction pointed out in previous studies~\cite{hiroaki-2015}.

	Next, we consider the $n_e\simeq3/4$ case. We have illustrated the results obtained from MC calculations for the magnitude of the peak in the spin structure factor at $\mathbf{q}=(0,\pi)$ while varying $T$ in Fig.~\ref{fig:temp-dep}(b). A magnetic phase transition similar to the case for $n_e\simeq1/4$ is also observed here. 
	The ground state at the lower temperature is a noncollinear flux state similar to the previous case. This similarity between the results for both cases can be ascribed to the symmetry of the band structure at one-quarter and three-quarter filling of itinerant electrons. In Fig.~\ref{fig:sq3d-band}(b) the electronic energy bands along symmetric points in the BZ are shown for a tight-binding model with a double-exchange term. The two values of $\mathbf{q}$ used are that of a flux state: $(0,\pi)$ and $(\pi,0)$. It is clear that the bands are symmetric for one-quarter and three-quarter filling of the electron density.
		
	\subsection{Role of frustration}
	
		\begin{figure}[htb]
			\centering
			\includegraphics[width=0.49\textwidth]{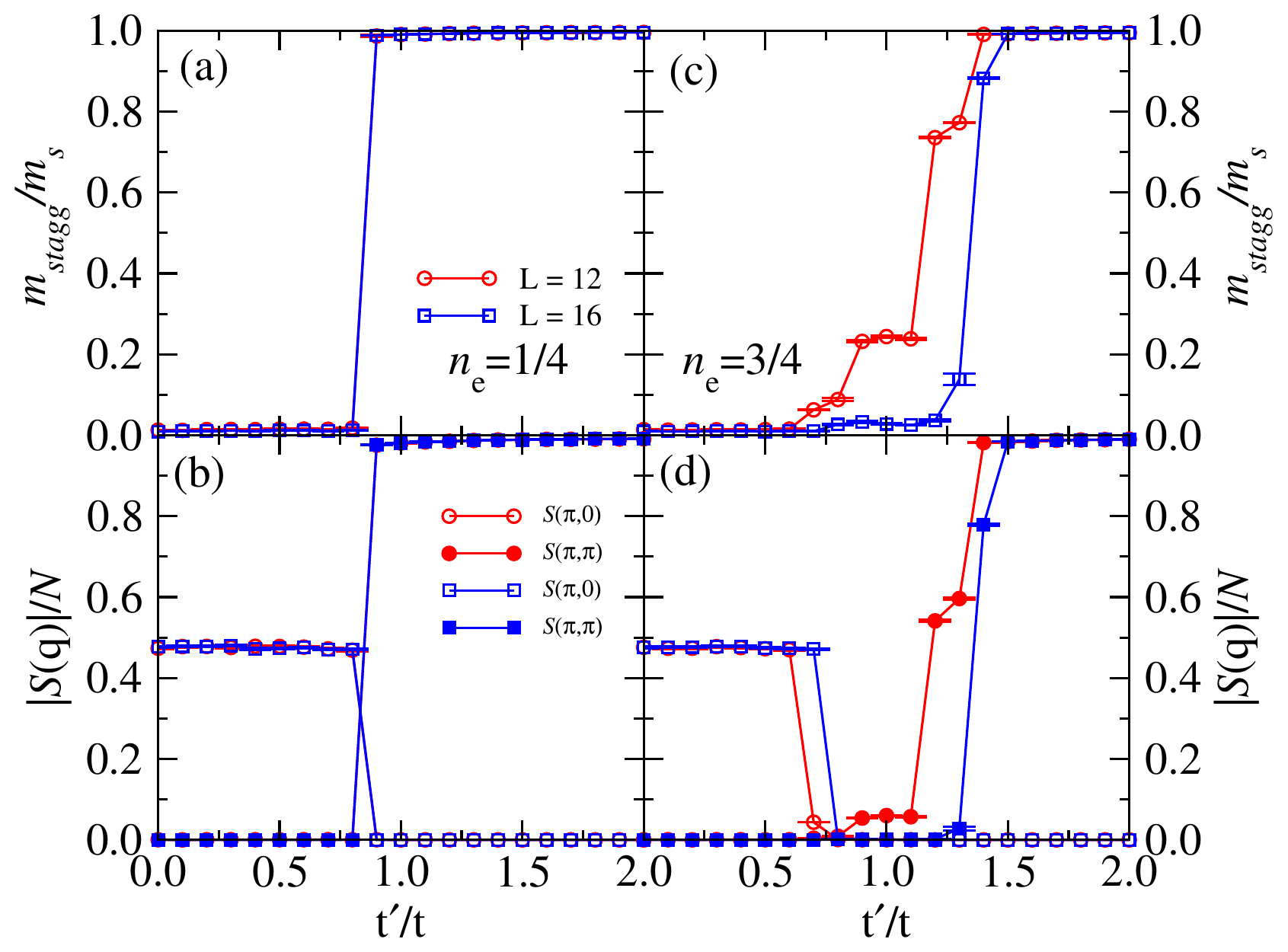}
			\caption{(Color online) The staggered magnetization per site while varying the frustration parameter $t'/t$ for (a) $n_e\simeq1/4$ and (c) $n_e\simeq3/4$. The frustration parameter $t'/t$ dependence of the magnitude of peaks in the spin structure factor at $\mathbf{q}=(\pi,0)$ and  $(\pi,\pi)$ for (b) $n_e\simeq1/4$ and (d) $n_e\simeq3/4$. Shown are the results for $12\times12$ and $16\times16$ lattice sizes while keeping the values of other parameters set to $J/t=0.1$, $J'/t=0.025$, $h^z/t=0$, $J_K/t=8.0$, and $T/t=0.005$.}
			\label{fig:tpr-dep}	
		\end{figure}
		
	\subsubsection{Frustration in electron hopping: $t'/t$}
		
In an accompanying paper, we have investigated the effects of Dzyaloshinskii-Moriya
interactions on the emergence and stability of noncoplanar configurations 	
of the local moments. In particular, we discovered that a novel canted flux state
is stabilized over an extended range of parameters (exchange interactions,
magnetic field) in the presence of DM interactions. In this work, 
we explore the role of frustration on the magnetic properties of the current model.  Both $t'$ and $J'$ induce frustration: the former in electron
hopping and the latter in exchange interactions between local moments. Since $t'$ and $J'$ are determined by the overlap of different orbitals across the diagonal
bonds, their ratio can, in principle, be different for different materials. In this section, we choose to vary
$t'$, keeping $J'$ constant.  With varying $t'$, 
the nature of electron hopping changes from being unfrustrated at $t'=0$ to being highly frustrated at $t' \gtrsim t$, whereas the degree of frustration in the direct exchange
between the local moments remains unaltered. We start with $n_e\simeq 1/4$; Figs.~\ref{fig:tpr-dep}(a) and \ref{fig:tpr-dep}(b) show the results of MC calculations for staggered magnetization and the magnitude of the peak in $S({\bf q})$ at $\mathbf{q}=(\pi,0)$ and  $(\pi,\pi)$,  respectively, while varying the hopping integral on the diagonal bonds $t'$. 
At $t'/t=0.0$, the ground state has almost zero staggered magnetization, and the static spin structure factor exhibits two sharp and equal-magnitude peaks at $(0,\pi)$ and $(\pi,0)$. Actually, this is the same flux state that we encountered in the last section. The in-plane components of the local moments are arranged in a pattern similar to a coplanar flux state.

The flux pattern consists of columnar AFM arrangements of the in-plane components
with simultaneous $(0,\pi)$ and $(\pi,0)$ ordering.
With increasing $t'/t$, the staggered magnetization remains constant at a value consistent with a noncollinear flux state up to $t'/t\approx0.8$, where there is a discontinuous transition to a state with a large value of staggered magnetization. There is a substantial increase in the magnitude of the peak in the static structure factor $S({\bf q})$ at $(\pi,\pi)$, indicating AFM ordering of the localized moments. The weight of $S({\bf q})$ is negligible for the flux ordering wave vectors $(\pi,0)$ and $(0,\pi)$, which specifies the breaking of in-plane flux ordering. Actually, this is a discontinuous spin-flop transition to a pure AFM state. The in-plane components are seen to evolve from a flux pattern at small $t'/t$ to staggered AFM order at large values of $t'/t$.

	\begin{figure}[htb]
		\centering
		\includegraphics[width=0.49\textwidth]{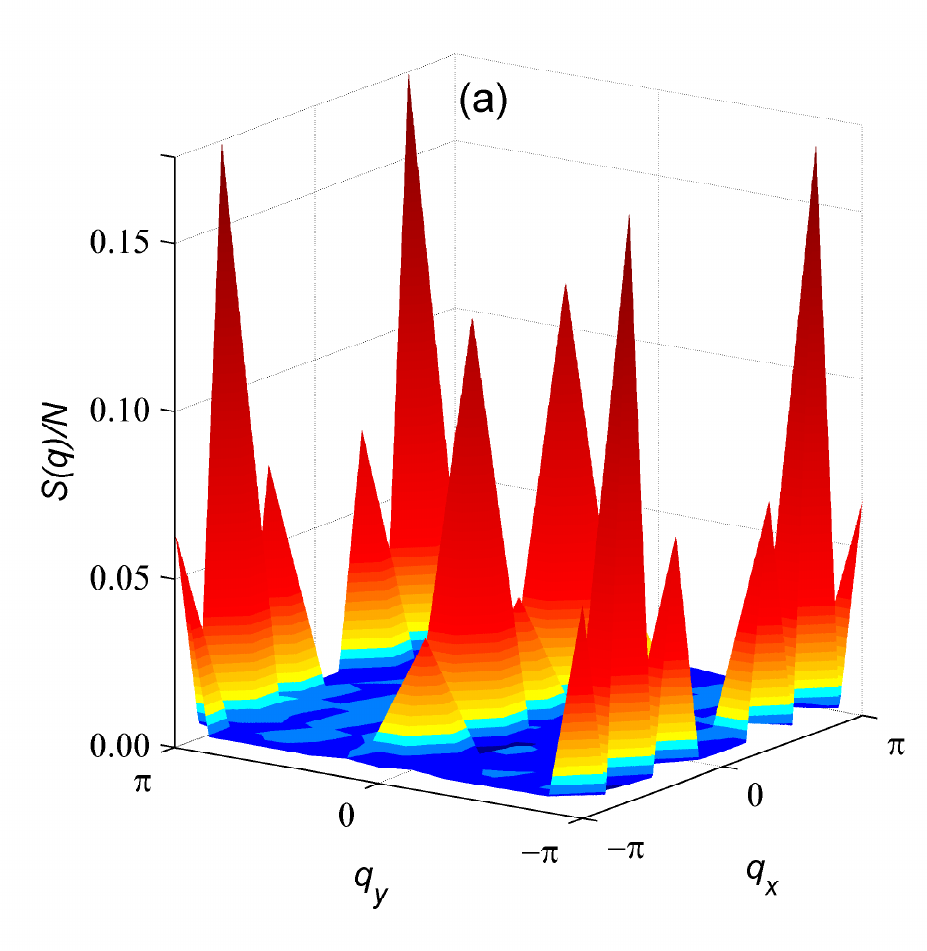}
		
		\includegraphics[width=0.49\textwidth]{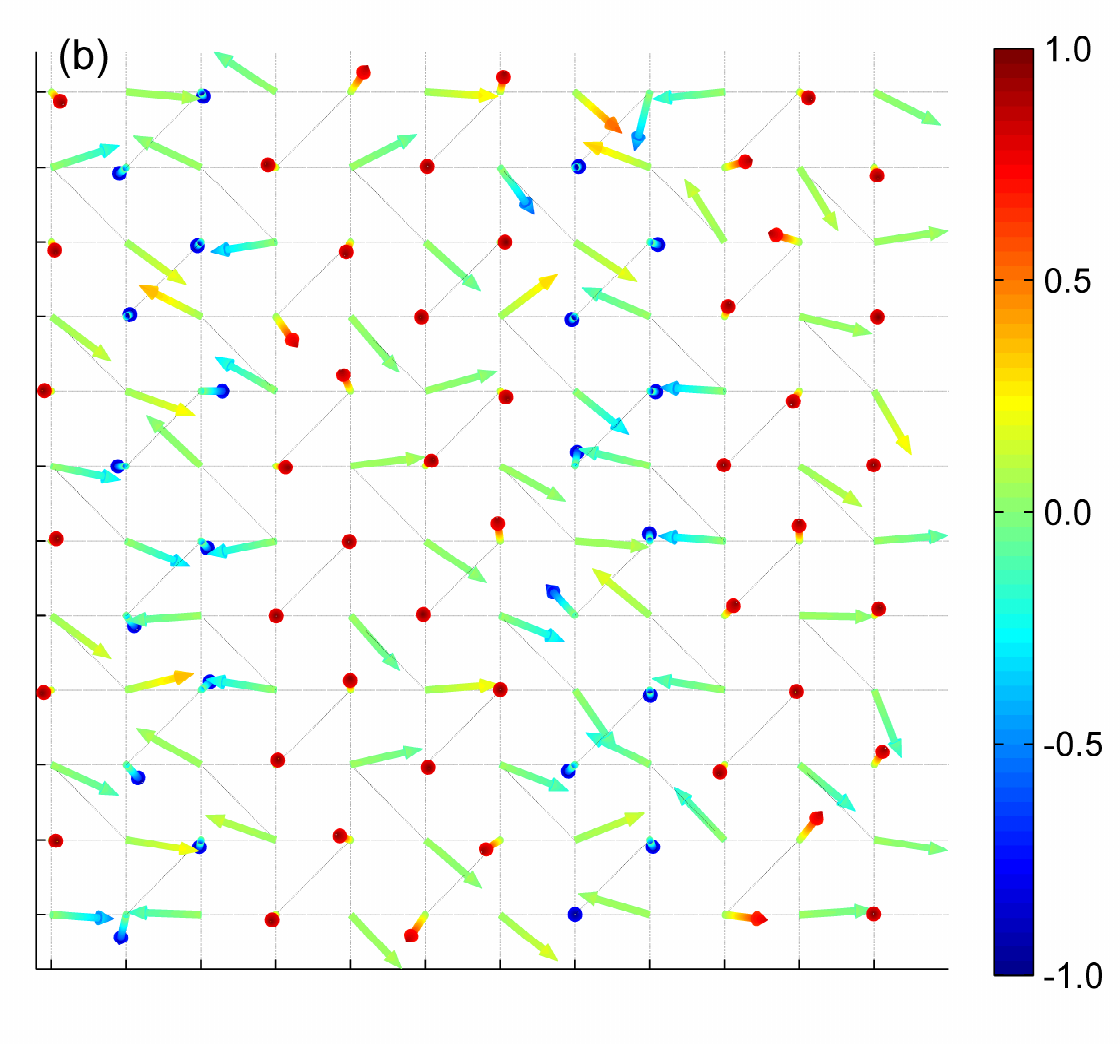}
		\caption{(Color online) (a) The static spin structure plotted against the momentum vectors and (b) the real spin configuration of localized spins obtained from MC calculations for the $12\times12$ lattice at $n_e\simeq3/4$, $t'/t=0.9$, $J/t=0.1$, $J'/t=0.025$, $h^z/t=0$, $J_K/t=8.0$, and $T/t=0.005$. The color bar represents the magnitude of out-of-plane components of localized spins.}
		\label{fig:3b4sq&confg}	
	\end{figure} 

On the other hand, the evolution of the magnetic ground state with changing 
$t'/t$ for $n_e\simeq3/4$ is markedly different, as seen in Figs.~\ref{fig:tpr-dep}(c) and \ref{fig:tpr-dep}(d). The nature of the magnetic ground state is similar to that for 
$n_e\simeq 1/4$, viz., the coplanar flux state at small $t'/t$ and the collinear AFM state at large $t'/t$, but the transition between the two is not a direct one.
Instead, there is an extended range of intermediate values of $t'/t$ for which the ground state exhibits neither flux nor staggered AFM ordering.
The values of the static structure factor is vanishingly small at 
${\bf q} =(\pi,\pi), (0,\pi)$, and $(\pi,0)$.
We observed sharp peaks in $S({\bf q})$ at $\mathbf{q}=(\pm2\pi/3,\pm\pi)$ and $(\pm\pi/3,0)$ but weak peaks at $(\pm\pi/3,\pm\pi)$ and $(\pi,\pi)$ [the magnitude of $S({\bf q})$ at these weak peaks is an order of magnitude
smaller than those for the sharp peaks] and even weaker satellites around 
${\bf q} =(\pm\pi/3,0)$ and $(0,0)$, pointing towards
a very weak long-range ordering [see Fig.~\ref{fig:3b4sq&confg}(a)]. A snapshot of the real-space configuration [shown in Fig.~\ref{fig:3b4sq&confg}(b)] of the local moments also does not exhibit any
obvious pattern of the spin orientations. The complete nature
of the magnetic ordering in this intriguing state and the reason for the anisotropy in the results for the two filling factors is currently being investigated.

\begin{figure}[htb]
	\centering
	\includegraphics[width=0.49\textwidth]{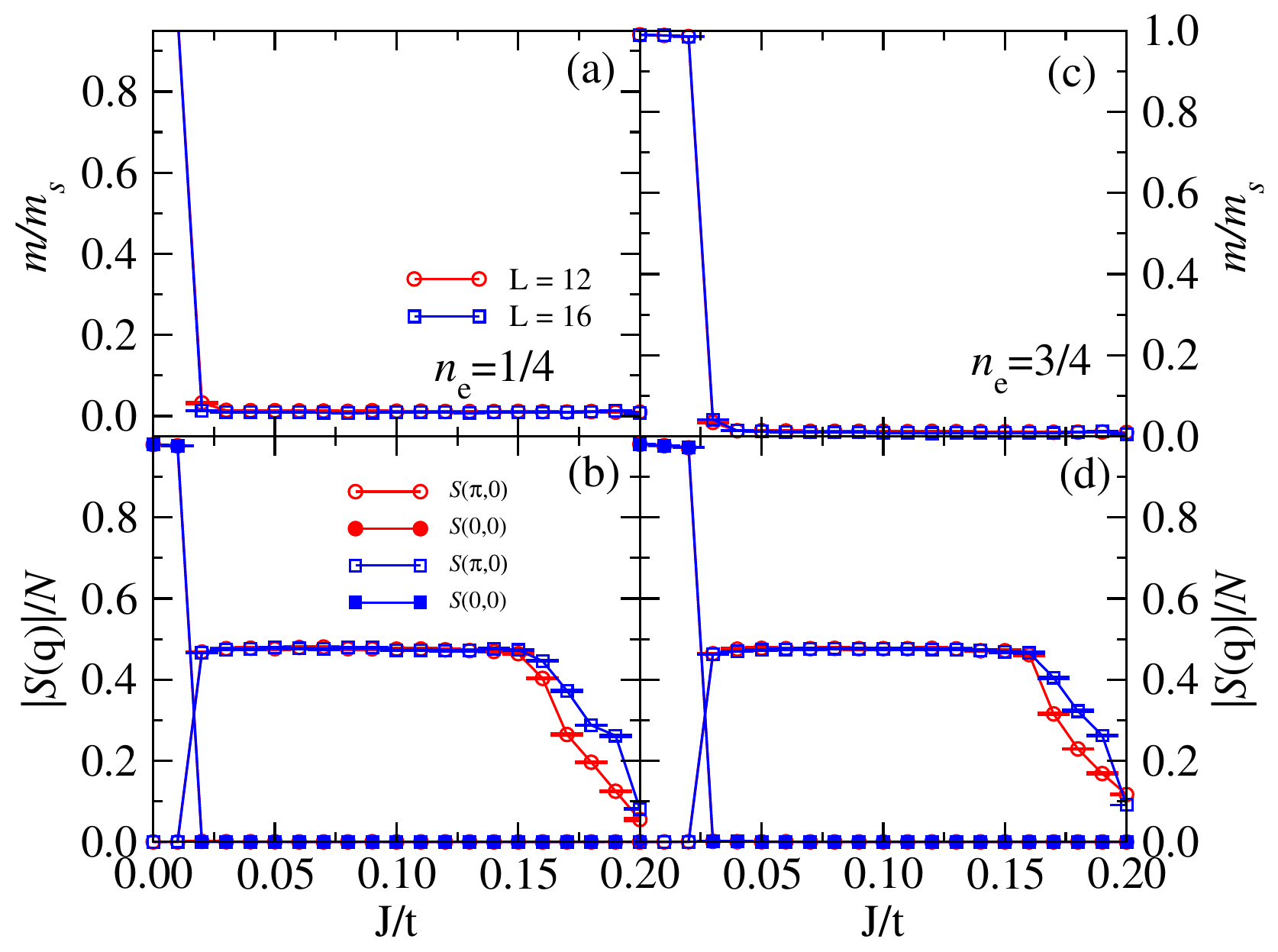}
	\caption{(Color online) The magnetization per site as a function of the frustration parameter $J/t$ for (a) $n_e\simeq1/4$ and (c) $n_e\simeq3/4$. The frustration parameter dependence $J/t$ of the magnitude in the spin static structure factor at $\mathbf{q}=(\pi,0)$ and $(0,0)$ for (b) $n_e\simeq1/4$ and (d) $n_e\simeq3/4$, respectively. The results are shown for $12\times12$ and $16\times16$ lattice sizes while keeping the values of other parameters set to $J'/t=0.025$, $t'/t=0.25$, $h^z/t=0$, $J_K/t=8.0$, and $T/t=0.005$.}
	\label{fig:jh-dep}	
\end{figure} 

\subsubsection{Frustration in exchange coupling: $J'/t$}

\begin{figure}[htb]
	\centering
	\includegraphics[width=0.49\textwidth]{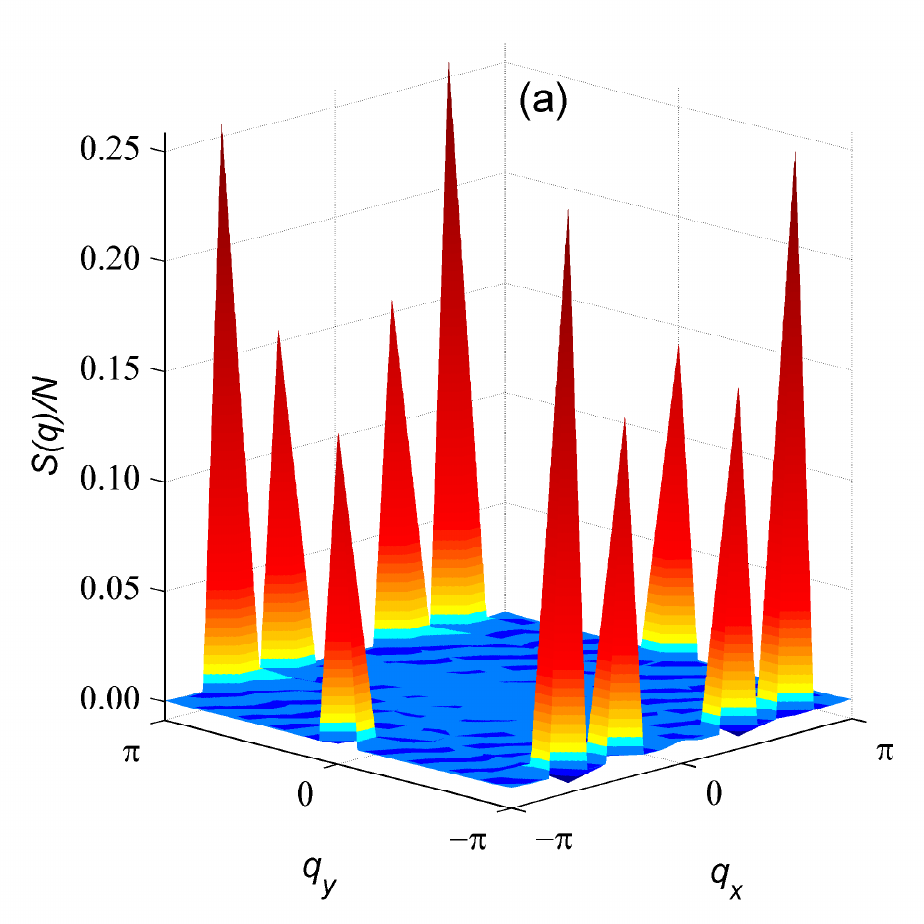}
	
	\includegraphics[width=0.49\textwidth]{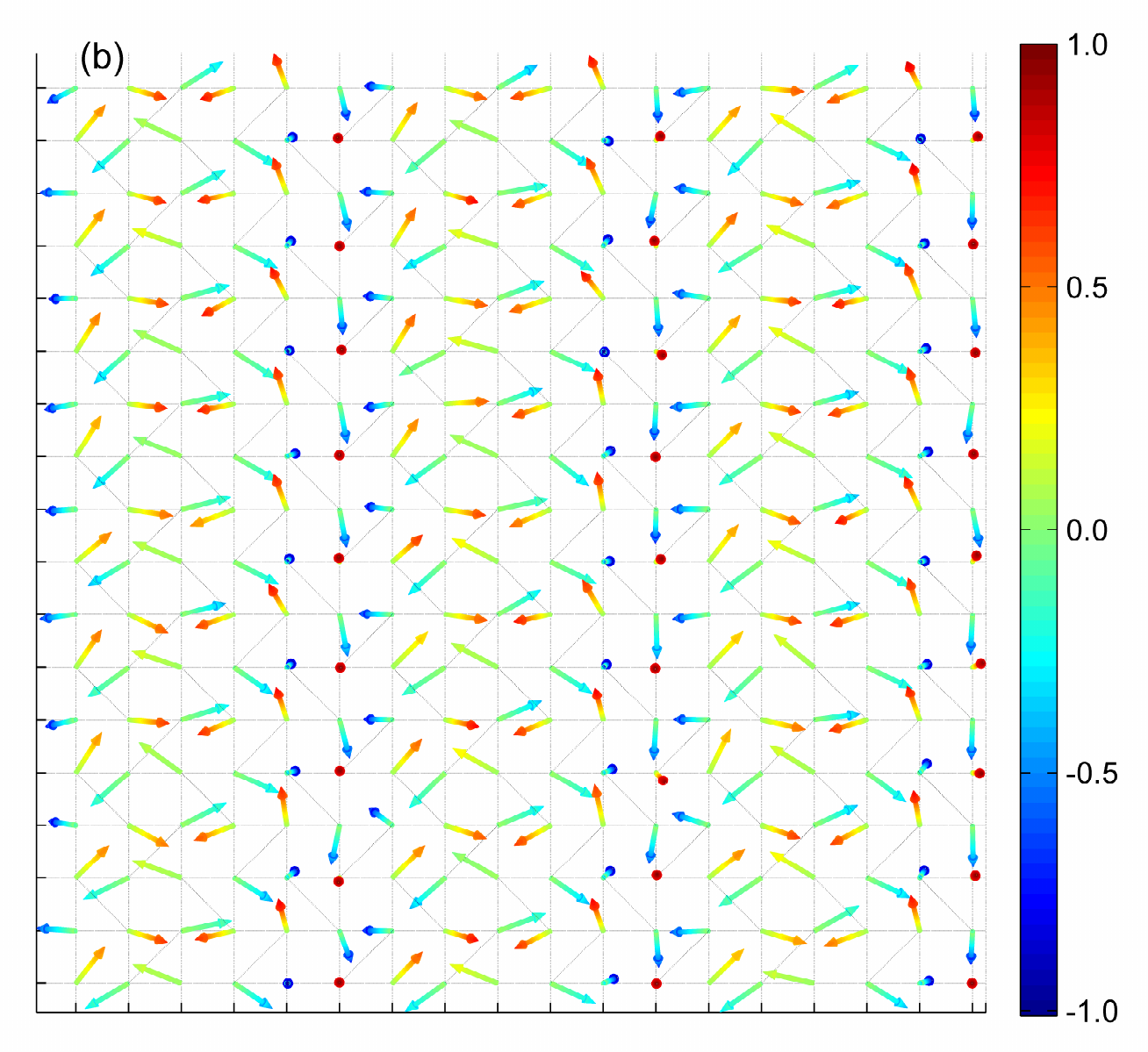}
	\caption{(Color online) (a) The static spin structure plotted as a function of $\mathbf{q}$ vectors and (b) a snapshot of the real-space configuration of localized spins obtained from MC calculations for an $18\times18$ lattice at $n_e\simeq1/4$, $J/t=0.19$, $J'/t=0.025$, $t'/t=0.25$, $h^z/t=0$, $J_K/t=8.0$, and $T/t=0.005$. The color bar represents the magnitude of out-of-plane components of localized spins.}
	\label{fig:confg}	
\end{figure}

In this section we fix the value of $t'/t$ and vary the ratio of $J/J'$. As mentioned earlier, this will vary the degree of frustration in the exchange coupling between the localized spins. We start with analyzing the results for $n_e\simeq1/4$. Figures~\ref{fig:jh-dep}(a) and \ref{fig:jh-dep}(b) show the evolution of uniform magnetization per site and magnitude of the peaks in $S({\bf q})$ at $(0,0)$ and $(\pi,0)$, respectively, as a function of $J/t$, keeping $J'/t$ constant. For a very small exchange interaction between NN localized spins the ground state is FM, as indicated by large uniform magnetization and a large value of the peak at $\mathbf{q}=(0,0)$. In the limit of strong $J_K$ the double-exchange mechanism stabilizes the FM ordering as there is a large kinetic energy gain if the spins on two sites are parallel. Around $J/t\simeq0.02$, there is a transition to a magnetic state for which the uniform magnetization drops to vanishingly small values and peaks in the structure factor appear at $(0,\pi)$ and $(\pi,0)$. This is the same noncollinear flux state that we observed earlier with two equal magnitude peaks at $\mathbf{q}=(0,\pi)$ and $(\pi,0)$ in $S({\bf q})$. This state is stabilized over a wide range of $J/t$. Further increasing the exchange interaction results in a ground state where the magnitude of the peak at $(\pi,0)$ decreases and peaks in $S(\mathbf{q})$ at $(\pm\pi/3,\pi)$ and $(\pm2\pi/3,\pi)$ appear [as shown in Fig.~\ref{fig:confg}(a)]. This indicates that large frustration in exchange coupling breaks the flux pattern between the localized spins. A snapshot of
the real-space configuration shows incommensurate magnetic ordering that consists of 
three unit-cell-wide stripes stacked parallel to the $y$ axis. Each stripe consists of 
antiferromagnetically ordered local moments, with a domain wall between adjacent stripes [see Fig.~\ref{fig:confg}(b)]. 

For $n_e\simeq3/4$, the results are qualitatively the same as shown in Figs.~\ref{fig:jh-dep}(c) and \ref{fig:jh-dep}(d), with the only difference being that the transition is shifted a bit to $J/t\simeq0.03$. The ground state evolves from a FM state at very small values of $J/t$ to a flux state in the intermediate regime to a state with peaks in the structure factor at $(\pm\pi/3,\pi)$ and $(\pm2\pi/3,\pi)$ at very large values of $J/t$.

	\subsection{Role of magnetic field}
	
\begin{figure}[htb]
		\centering
		\includegraphics[width=0.49\textwidth]{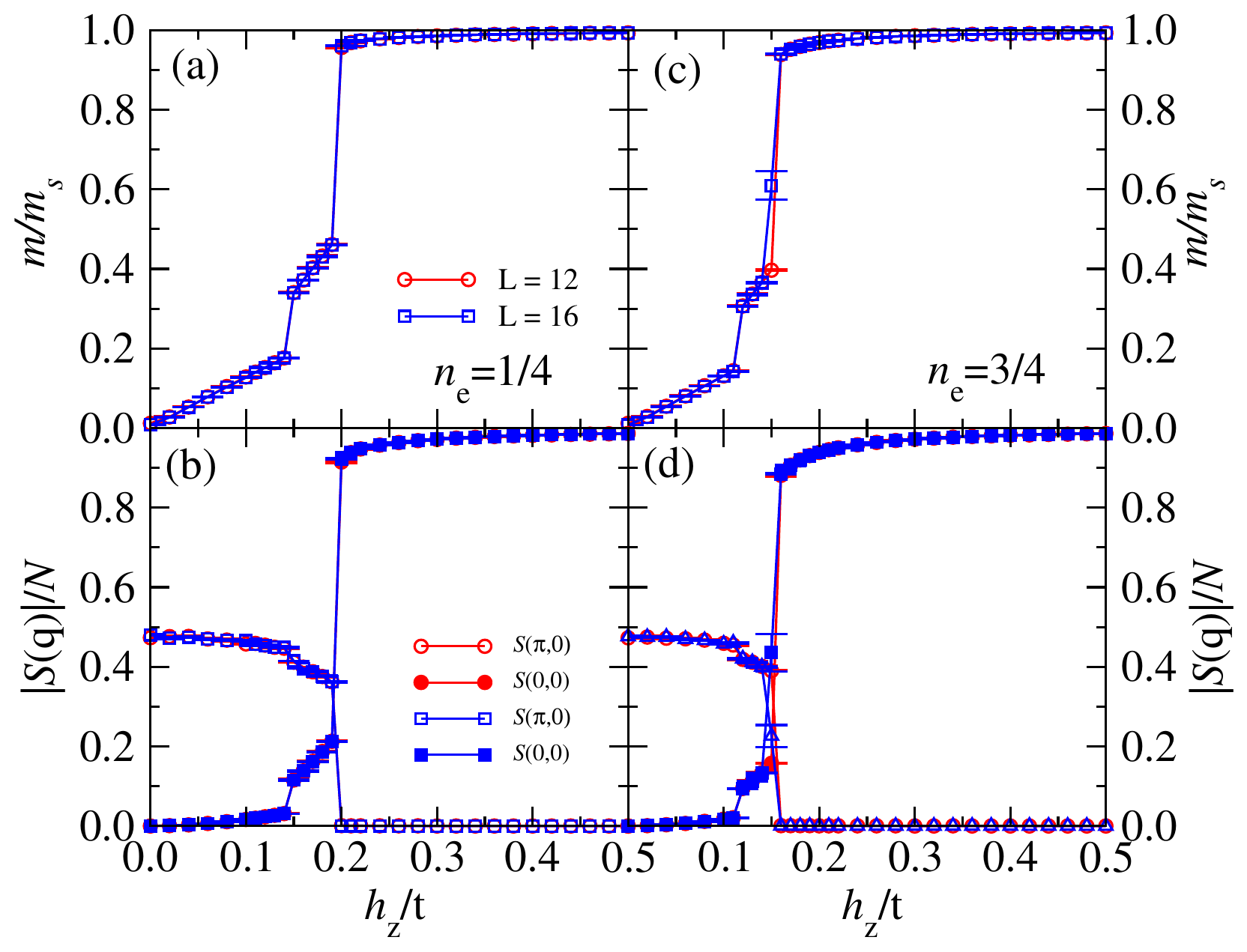}
		\caption{(Color online) The magnetic field dependence of magnetization per site for (a) $n_e\simeq1/4$ and (c) $n_e\simeq3/4$. The magnitude of the peaks in the static spin structure factor at  $\mathbf{q}=(\pi,0)$ and $(0,0)$ plotted while varying the strength of an external magnetic field for (b) $n_e\simeq1/4$ and (d) $ne\simeq3/4$. The results are shown for $12\times12$ and $16\times16$ lattice sizes while keeping the values of other parameters set to $t'/t=0.25$, $J/t=0.1$, $J'/t=0.025$, $J_K/t=8.0$, and $T/t=0.005$.}
		\label{fig:mag-field}	
	\end{figure}

	Finally, we investigate the role of an external magnetic field in the current model at both one-quarter and three-quarter fillings of electrons. Applying an external field is the simplest
and most direct way of controlling the magnetic character of a system. Our goal is to explore
the tunability of different magnetic states by applying a static, uniform, longitudinal external field. The canonical (purely magnetic) Shastry-Sutherland model exhibits a sequence of unique magnetization plateaus in an applied magnetic field. However, the strong coupling between local moments and 
itinerant electrons suppresses the magnetization plateaus completely in the 
present SS-KLM. 
For $n_e\simeq1/4$ the magnetic field dependence of uniform magnetization and the magnitude of the peaks in $S({\bf q})$ at $(0,0)$ and $(\pi,0)$ is shown in Figs.~\ref{fig:mag-field}(a) and \ref{fig:mag-field}(b), respectively. At $h^z=0.0$, the magnetic ground state is a flux state, which we have already discussed in detail. The uniform magnetization is zero at zero field because the direct exchange between the local
moments is AFM in nature. 
The magnetization $m/m_s$ increases monotonically up to 
$h_z=0.20$, where there is an abrupt jump in its value from $\approx 0.4$ to $\approx 0.95$,
marking a field-driven discontinuous transition. 
The transition marks the breaking of the flux pattern
of the in-plane components of the local moments. This is confirmed by the behavior of the 
static structure factor at $(0,0)$ and $(\pi,0)$. At the transition, the peak at $(\pi,0)$ is 
completely suppressed, whereas the $(0,0)$ peak (proportional to the square of the
uniform magnetization) exhibits a discontinuous increase in its value. With further increase
in magnetic field, the system approaches full polarization asymptotically. Once again, the field
dependence of the magnetic ground state is qualitatively similar for $n_e\simeq3/4$, confirming
the particle-hole symmetry of the system.

	\section{Summary}\label{sec:summary}
	
	To summarize we have studied in detail the magnetic properties of the SS-KLM at filling factors of itinerant electrons $n_e=1/4$ and $3/4$ at which the ground state is metallic. We have found that a noncollinear flux state is stabilized over a large range of parameters at both densities. Interestingly, in contrast to insulating ground states, a Dzyaloshinskii-Moriya interaction is not essential for stabilizing noncollinear spin textures. The noncollinearity of the magnetic ground state can be suppressed via a discontinuous phase transition by external fields such as temperature, by a static, uniform longitudinal magnetic field, and by increasing the magnitude of frustration in electronic and exchange coupling components of the Hamiltonian. In fact, tuning the strength of diagonal hopping (which is responsible for frustration in the electronic component) drives the magnetic ground state from a flux phase at small frustration to a pure AFM state at large frustration. For three-quarter filling of itinerant electrons the transition from flux to AFM ground state is not direct and is accompanied by an intermediate state. We plan to investigate this intriguing state in the near future. On the other hand, the ground state for small exchange coupling is FM and changes to a flux state at intermediate values to a state with incommensurate magnetic ordering at large values of exchange coupling.
	 
    \begin{acknowledgements}
	
    The work was partially supported by Grant No. MOE2014-T2-2-112 from the Ministry of Education, Singapore. We also acknowledge the use of HPC cluster at NSSC, Singapore.
    \end{acknowledgements}

    \bibliographystyle{apsrev4-1}
    \bibliography{bibfile}

\begin{thebibliography}{49}%
\makeatletter
\providecommand \@ifxundefined [1]{%
 \@ifx{#1\undefined}
}%
\providecommand \@ifnum [1]{%
 \ifnum #1\expandafter \@firstoftwo
 \else \expandafter \@secondoftwo
 \fi
}%
\providecommand \@ifx [1]{%
 \ifx #1\expandafter \@firstoftwo
 \else \expandafter \@secondoftwo
 \fi
}%
\providecommand \natexlab [1]{#1}%
\providecommand \enquote  [1]{``#1''}%
\providecommand \bibnamefont  [1]{#1}%
\providecommand \bibfnamefont [1]{#1}%
\providecommand \citenamefont [1]{#1}%
\providecommand \href@noop [0]{\@secondoftwo}%
\providecommand \href [0]{\begingroup \@sanitize@url \@href}%
\providecommand \@href[1]{\@@startlink{#1}\@@href}%
\providecommand \@@href[1]{\endgroup#1\@@endlink}%
\providecommand \@sanitize@url [0]{\catcode `\\12\catcode `\$12\catcode
  `\&12\catcode `\#12\catcode `\^12\catcode `\_12\catcode `\%12\relax}%
\providecommand \@@startlink[1]{}%
\providecommand \@@endlink[0]{}%
\providecommand \url  [0]{\begingroup\@sanitize@url \@url }%
\providecommand \@url [1]{\endgroup\@href {#1}{\urlprefix }}%
\providecommand \urlprefix  [0]{URL }%
\providecommand \Eprint [0]{\href }%
\providecommand \doibase [0]{http://dx.doi.org/}%
\providecommand \selectlanguage [0]{\@gobble}%
\providecommand \bibinfo  [0]{\@secondoftwo}%
\providecommand \bibfield  [0]{\@secondoftwo}%
\providecommand \translation [1]{[#1]}%
\providecommand \BibitemOpen [0]{}%
\providecommand \bibitemStop [0]{}%
\providecommand \bibitemNoStop [0]{.\EOS\space}%
\providecommand \EOS [0]{\spacefactor3000\relax}%
\providecommand \BibitemShut  [1]{\csname bibitem#1\endcsname}%
\let\auto@bib@innerbib\@empty
\bibitem [{\citenamefont {Zener}(1951)}]{zener-1951}%
  \BibitemOpen
  \bibfield  {author} {\bibinfo {author} {\bibfnamefont {C.}~\bibnamefont
  {Zener}},\ }\href {\doibase 10.1103/PhysRev.82.403} {\bibfield  {journal}
  {\bibinfo  {journal} {Phys. Rev.}\ }\textbf {\bibinfo {volume} {82}},\
  \bibinfo {pages} {403} (\bibinfo {year} {1951})}\BibitemShut {NoStop}%
\bibitem [{\citenamefont {Anderson}\ and\ \citenamefont
  {Hasegawa}(1955)}]{anderson-1955}%
  \BibitemOpen
  \bibfield  {author} {\bibinfo {author} {\bibfnamefont {P.~W.}\ \bibnamefont
  {Anderson}}\ and\ \bibinfo {author} {\bibfnamefont {H.}~\bibnamefont
  {Hasegawa}},\ }\href {\doibase 10.1103/PhysRev.100.675} {\bibfield  {journal}
  {\bibinfo  {journal} {Phys. Rev.}\ }\textbf {\bibinfo {volume} {100}},\
  \bibinfo {pages} {675} (\bibinfo {year} {1955})}\BibitemShut {NoStop}%
\bibitem [{\citenamefont {Furukawa}(1995)}]{furukawa-1995}%
  \BibitemOpen
  \bibfield  {author} {\bibinfo {author} {\bibfnamefont {N.}~\bibnamefont
  {Furukawa}},\ }\href {\doibase 10.1143/JPSJ.64.2734} {\bibfield  {journal}
  {\bibinfo  {journal} {J. Phys. Soc. Jpn.}\ }\textbf {\bibinfo {volume}
  {64}},\ \bibinfo {pages} {2734} (\bibinfo {year} {1995})}\BibitemShut
  {NoStop}%
\bibitem [{\citenamefont {Yunoki}\ \emph {et~al.}(1998)\citenamefont {Yunoki},
  \citenamefont {Hu}, \citenamefont {Malvezzi}, \citenamefont {Moreo},
  \citenamefont {Furukawa},\ and\ \citenamefont {Dagotto}}]{yunoki-1998}%
  \BibitemOpen
  \bibfield  {author} {\bibinfo {author} {\bibfnamefont {S.}~\bibnamefont
  {Yunoki}}, \bibinfo {author} {\bibfnamefont {J.}~\bibnamefont {Hu}}, \bibinfo
  {author} {\bibfnamefont {A.~L.}\ \bibnamefont {Malvezzi}}, \bibinfo {author}
  {\bibfnamefont {A.}~\bibnamefont {Moreo}}, \bibinfo {author} {\bibfnamefont
  {N.}~\bibnamefont {Furukawa}}, \ and\ \bibinfo {author} {\bibfnamefont
  {E.}~\bibnamefont {Dagotto}},\ }\href {\doibase 10.1103/PhysRevLett.80.845}
  {\bibfield  {journal} {\bibinfo  {journal} {Phys. Rev. Lett.}\ }\textbf
  {\bibinfo {volume} {80}},\ \bibinfo {pages} {845} (\bibinfo {year}
  {1998})}\BibitemShut {NoStop}%
\bibitem [{\citenamefont {Kubo}\ and\ \citenamefont {Ohata}(1972)}]{kubo-1972}%
  \BibitemOpen
  \bibfield  {author} {\bibinfo {author} {\bibfnamefont {K.}~\bibnamefont
  {Kubo}}\ and\ \bibinfo {author} {\bibfnamefont {N.}~\bibnamefont {Ohata}},\
  }\href {\doibase 10.1143/JPSJ.33.21} {\bibfield  {journal} {\bibinfo
  {journal} {J. Phys. Soc. Jpn.}\ }\textbf {\bibinfo {volume} {33}},\ \bibinfo
  {pages} {21} (\bibinfo {year} {1972})}\BibitemShut {NoStop}%
\bibitem [{\citenamefont {Grohol}\ \emph {et~al.}(2005)\citenamefont {Grohol},
  \citenamefont {Matan}, \citenamefont {Cho}, \citenamefont {Lee},
  \citenamefont {Lynn}, \citenamefont {Nocera},\ and\ \citenamefont
  {Lee}}]{grohol-2005}%
  \BibitemOpen
  \bibfield  {author} {\bibinfo {author} {\bibfnamefont {D.}~\bibnamefont
  {Grohol}}, \bibinfo {author} {\bibfnamefont {K.}~\bibnamefont {Matan}},
  \bibinfo {author} {\bibfnamefont {J.-H.}\ \bibnamefont {Cho}}, \bibinfo
  {author} {\bibfnamefont {S.-H.}\ \bibnamefont {Lee}}, \bibinfo {author}
  {\bibfnamefont {J.~W.}\ \bibnamefont {Lynn}}, \bibinfo {author}
  {\bibfnamefont {D.~G.}\ \bibnamefont {Nocera}}, \ and\ \bibinfo {author}
  {\bibfnamefont {Y.~S.}\ \bibnamefont {Lee}},\ }\href {\doibase
  10.1038/nmat1353} {\bibfield  {journal} {\bibinfo  {journal} {Nat. Mater.}\
  }\textbf {\bibinfo {volume} {4}},\ \bibinfo {pages} {323} (\bibinfo {year}
  {2005})}\BibitemShut {NoStop}%
\bibitem [{\citenamefont {Balicas}\ \emph {et~al.}(2011)\citenamefont
  {Balicas}, \citenamefont {Nakatsuji}, \citenamefont {Machida},\ and\
  \citenamefont {Onoda}}]{balicas-2011}%
  \BibitemOpen
  \bibfield  {author} {\bibinfo {author} {\bibfnamefont {L.}~\bibnamefont
  {Balicas}}, \bibinfo {author} {\bibfnamefont {S.}~\bibnamefont {Nakatsuji}},
  \bibinfo {author} {\bibfnamefont {Y.}~\bibnamefont {Machida}}, \ and\
  \bibinfo {author} {\bibfnamefont {S.}~\bibnamefont {Onoda}},\ }\href
  {\doibase 10.1103/PhysRevLett.106.217204} {\bibfield  {journal} {\bibinfo
  {journal} {Phys. Rev. Lett.}\ }\textbf {\bibinfo {volume} {106}},\ \bibinfo
  {pages} {217204} (\bibinfo {year} {2011})}\BibitemShut {NoStop}%
\bibitem [{\citenamefont {Wen}\ \emph {et~al.}(1989)\citenamefont {Wen},
  \citenamefont {Wilczek},\ and\ \citenamefont {Zee}}]{wen-1989}%
  \BibitemOpen
  \bibfield  {author} {\bibinfo {author} {\bibfnamefont {X.~G.}\ \bibnamefont
  {Wen}}, \bibinfo {author} {\bibfnamefont {F.}~\bibnamefont {Wilczek}}, \ and\
  \bibinfo {author} {\bibfnamefont {A.}~\bibnamefont {Zee}},\ }\href {\doibase
  10.1103/PhysRevB.39.11413} {\bibfield  {journal} {\bibinfo  {journal} {Phys.
  Rev. B}\ }\textbf {\bibinfo {volume} {39}},\ \bibinfo {pages} {11413}
  (\bibinfo {year} {1989})}\BibitemShut {NoStop}%
\bibitem [{\citenamefont {Wen}(1991)}]{wen-1991}%
  \BibitemOpen
  \bibfield  {author} {\bibinfo {author} {\bibfnamefont {X.~G.}\ \bibnamefont
  {Wen}},\ }\href {\doibase 10.1103/PhysRevB.43.11025} {\bibfield  {journal}
  {\bibinfo  {journal} {Phys. Rev. B}\ }\textbf {\bibinfo {volume} {43}},\
  \bibinfo {pages} {11025} (\bibinfo {year} {1991})}\BibitemShut {NoStop}%
\bibitem [{\citenamefont {Laughlin}\ and\ \citenamefont
  {Zou}(1990)}]{laughlin-1990}%
  \BibitemOpen
  \bibfield  {author} {\bibinfo {author} {\bibfnamefont {R.~B.}\ \bibnamefont
  {Laughlin}}\ and\ \bibinfo {author} {\bibfnamefont {Z.}~\bibnamefont {Zou}},\
  }\href {\doibase 10.1103/PhysRevB.41.664} {\bibfield  {journal} {\bibinfo
  {journal} {Phys. Rev. B}\ }\textbf {\bibinfo {volume} {41}},\ \bibinfo
  {pages} {664} (\bibinfo {year} {1990})}\BibitemShut {NoStop}%
\bibitem [{\citenamefont {Udagawa}\ and\ \citenamefont
  {Moessner}(2013)}]{udagawa-2013}%
  \BibitemOpen
  \bibfield  {author} {\bibinfo {author} {\bibfnamefont {M.}~\bibnamefont
  {Udagawa}}\ and\ \bibinfo {author} {\bibfnamefont {R.}~\bibnamefont
  {Moessner}},\ }\href {\doibase 10.1103/PhysRevLett.111.036602} {\bibfield
  {journal} {\bibinfo  {journal} {Phys. Rev. Lett.}\ }\textbf {\bibinfo
  {volume} {111}},\ \bibinfo {pages} {036602} (\bibinfo {year}
  {2013})}\BibitemShut {NoStop}%
\bibitem [{\citenamefont {Boldrin}\ and\ \citenamefont
  {Wills}()}]{boldrin-2012}%
  \BibitemOpen
  \bibfield  {author} {\bibinfo {author} {\bibfnamefont {D.}~\bibnamefont
  {Boldrin}}\ and\ \bibinfo {author} {\bibfnamefont {A.}~\bibnamefont
  {Wills}},\ }\href {\doibase 10.1155/2012/615295} {\bibfield  {journal}
  {\bibinfo  {journal} {Adv. Condens. Matter Phys.}\ }\textbf {\bibinfo
  {volume} {2012}},\ \bibinfo {pages} {615295}}\BibitemShut {NoStop}%
\bibitem [{\citenamefont {Taguchi}\ \emph {et~al.}(2001)\citenamefont
  {Taguchi}, \citenamefont {Oohara}, \citenamefont {Yoshizawa}, \citenamefont
  {Nagaosa},\ and\ \citenamefont {Tokura}}]{taguchi-2001}%
  \BibitemOpen
  \bibfield  {author} {\bibinfo {author} {\bibfnamefont {Y.}~\bibnamefont
  {Taguchi}}, \bibinfo {author} {\bibfnamefont {Y.}~\bibnamefont {Oohara}},
  \bibinfo {author} {\bibfnamefont {H.}~\bibnamefont {Yoshizawa}}, \bibinfo
  {author} {\bibfnamefont {N.}~\bibnamefont {Nagaosa}}, \ and\ \bibinfo
  {author} {\bibfnamefont {Y.}~\bibnamefont {Tokura}},\ }\href {\doibase
  10.1126/science.1058161} {\bibfield  {journal} {\bibinfo  {journal}
  {Science}\ }\textbf {\bibinfo {volume} {291}},\ \bibinfo {pages} {2573}
  (\bibinfo {year} {2001})}\BibitemShut {NoStop}%
\bibitem [{\citenamefont {Nagaosa}(2001)}]{nagaosa-2001}%
  \BibitemOpen
  \bibfield  {author} {\bibinfo {author} {\bibfnamefont {N.}~\bibnamefont
  {Nagaosa}},\ }\href {\doibase
  http://dx.doi.org/10.1016/S0921-5107(01)00570-0} {\bibfield  {journal}
  {\bibinfo  {journal} {Mater. Sci. Engi. B}\ }\textbf {\bibinfo {volume}
  {84}},\ \bibinfo {pages} {58 } (\bibinfo {year} {2001})}\BibitemShut
  {NoStop}%
\bibitem [{\citenamefont {Machida}\ \emph {et~al.}(2010)\citenamefont
  {Machida}, \citenamefont {Nakatsuji}, \citenamefont {Onoda}, \citenamefont
  {Tayama},\ and\ \citenamefont {Sakakibara}}]{machida-2010}%
  \BibitemOpen
  \bibfield  {author} {\bibinfo {author} {\bibfnamefont {Y.}~\bibnamefont
  {Machida}}, \bibinfo {author} {\bibfnamefont {S.}~\bibnamefont {Nakatsuji}},
  \bibinfo {author} {\bibfnamefont {S.}~\bibnamefont {Onoda}}, \bibinfo
  {author} {\bibfnamefont {T.}~\bibnamefont {Tayama}}, \ and\ \bibinfo {author}
  {\bibfnamefont {T.}~\bibnamefont {Sakakibara}},\ }\href {\doibase
  10.1038/nature08680} {\bibfield  {journal} {\bibinfo  {journal} {Nature
  (London)}\ }\textbf {\bibinfo {volume} {463}},\ \bibinfo {pages} {210}
  (\bibinfo {year} {2010})}\BibitemShut {NoStop}%
\bibitem [{\citenamefont {Machida}\ \emph {et~al.}(2007)\citenamefont
  {Machida}, \citenamefont {Nakatsuji}, \citenamefont {Maeno}, \citenamefont
  {Tayama}, \citenamefont {Sakakibara},\ and\ \citenamefont
  {Onoda}}]{machida-2007}%
  \BibitemOpen
  \bibfield  {author} {\bibinfo {author} {\bibfnamefont {Y.}~\bibnamefont
  {Machida}}, \bibinfo {author} {\bibfnamefont {S.}~\bibnamefont {Nakatsuji}},
  \bibinfo {author} {\bibfnamefont {Y.}~\bibnamefont {Maeno}}, \bibinfo
  {author} {\bibfnamefont {T.}~\bibnamefont {Tayama}}, \bibinfo {author}
  {\bibfnamefont {T.}~\bibnamefont {Sakakibara}}, \ and\ \bibinfo {author}
  {\bibfnamefont {S.}~\bibnamefont {Onoda}},\ }\href {\doibase
  10.1103/PhysRevLett.98.057203} {\bibfield  {journal} {\bibinfo  {journal}
  {Phys. Rev. Lett.}\ }\textbf {\bibinfo {volume} {98}},\ \bibinfo {pages}
  {057203} (\bibinfo {year} {2007})}\BibitemShut {NoStop}%
\bibitem [{\citenamefont {Ye}\ \emph {et~al.}(1999)\citenamefont {Ye},
  \citenamefont {Kim}, \citenamefont {Millis}, \citenamefont {Shraiman},
  \citenamefont {Majumdar},\ and\ \citenamefont {Te\ifmmode \check{s}\else
  \v{s}\fi{}anovi\ifmmode~\acute{c}\else \'{c}\fi{}}}]{jinwu-1999}%
  \BibitemOpen
  \bibfield  {author} {\bibinfo {author} {\bibfnamefont {J.}~\bibnamefont
  {Ye}}, \bibinfo {author} {\bibfnamefont {Y.~B.}\ \bibnamefont {Kim}},
  \bibinfo {author} {\bibfnamefont {A.~J.}\ \bibnamefont {Millis}}, \bibinfo
  {author} {\bibfnamefont {B.~I.}\ \bibnamefont {Shraiman}}, \bibinfo {author}
  {\bibfnamefont {P.}~\bibnamefont {Majumdar}}, \ and\ \bibinfo {author}
  {\bibfnamefont {Z.}~\bibnamefont {Te\ifmmode \check{s}\else
  \v{s}\fi{}anovi\ifmmode~\acute{c}\else \'{c}\fi{}}},\ }\href {\doibase
  10.1103/PhysRevLett.83.3737} {\bibfield  {journal} {\bibinfo  {journal}
  {Phys. Rev. Lett.}\ }\textbf {\bibinfo {volume} {83}},\ \bibinfo {pages}
  {3737} (\bibinfo {year} {1999})}\BibitemShut {NoStop}%
\bibitem [{\citenamefont {Berry}(1984)}]{berry-1984}%
  \BibitemOpen
  \bibfield  {author} {\bibinfo {author} {\bibfnamefont {M.~V.}\ \bibnamefont
  {Berry}},\ }\href {\doibase 10.1098/rspa.1984.0023} {\bibfield  {journal}
  {\bibinfo  {journal} {Proc. R. Soc. London, Ser. A}\ }\textbf {\bibinfo
  {volume} {392}},\ \bibinfo {pages} {45} (\bibinfo {year} {1984})}\BibitemShut
  {NoStop}%
\bibitem [{\citenamefont {Loss}\ and\ \citenamefont
  {Goldbart}(1992)}]{daniel-1992}%
  \BibitemOpen
  \bibfield  {author} {\bibinfo {author} {\bibfnamefont {D.}~\bibnamefont
  {Loss}}\ and\ \bibinfo {author} {\bibfnamefont {P.~M.}\ \bibnamefont
  {Goldbart}},\ }\href {\doibase 10.1103/PhysRevB.45.13544} {\bibfield
  {journal} {\bibinfo  {journal} {Phys. Rev. B}\ }\textbf {\bibinfo {volume}
  {45}},\ \bibinfo {pages} {13544} (\bibinfo {year} {1992})}\BibitemShut
  {NoStop}%
\bibitem [{\citenamefont {Ohgushi}\ \emph {et~al.}(2000)\citenamefont
  {Ohgushi}, \citenamefont {Murakami},\ and\ \citenamefont
  {Nagaosa}}]{kenya-2000}%
  \BibitemOpen
  \bibfield  {author} {\bibinfo {author} {\bibfnamefont {K.}~\bibnamefont
  {Ohgushi}}, \bibinfo {author} {\bibfnamefont {S.}~\bibnamefont {Murakami}}, \
  and\ \bibinfo {author} {\bibfnamefont {N.}~\bibnamefont {Nagaosa}},\ }\href
  {\doibase 10.1103/PhysRevB.62.R6065} {\bibfield  {journal} {\bibinfo
  {journal} {Phys. Rev. B}\ }\textbf {\bibinfo {volume} {62}},\ \bibinfo
  {pages} {R6065} (\bibinfo {year} {2000})}\BibitemShut {NoStop}%
\bibitem [{\citenamefont {Kato}\ \emph {et~al.}(2010)\citenamefont {Kato},
  \citenamefont {Martin},\ and\ \citenamefont {Batista}}]{yasu-2010}%
  \BibitemOpen
  \bibfield  {author} {\bibinfo {author} {\bibfnamefont {Y.}~\bibnamefont
  {Kato}}, \bibinfo {author} {\bibfnamefont {I.}~\bibnamefont {Martin}}, \ and\
  \bibinfo {author} {\bibfnamefont {C.~D.}\ \bibnamefont {Batista}},\ }\href
  {\doibase 10.1103/PhysRevLett.105.266405} {\bibfield  {journal} {\bibinfo
  {journal} {Phys. Rev. Lett.}\ }\textbf {\bibinfo {volume} {105}},\ \bibinfo
  {pages} {266405} (\bibinfo {year} {2010})}\BibitemShut {NoStop}%
\bibitem [{\citenamefont {Barros}\ and\ \citenamefont
  {Kato}(2013)}]{kipton-2013}%
  \BibitemOpen
  \bibfield  {author} {\bibinfo {author} {\bibfnamefont {K.}~\bibnamefont
  {Barros}}\ and\ \bibinfo {author} {\bibfnamefont {Y.}~\bibnamefont {Kato}},\
  }\href {\doibase 10.1103/PhysRevB.88.235101} {\bibfield  {journal} {\bibinfo
  {journal} {Phys. Rev. B}\ }\textbf {\bibinfo {volume} {88}},\ \bibinfo
  {pages} {235101} (\bibinfo {year} {2013})}\BibitemShut {NoStop}%
\bibitem [{\citenamefont {Rahmani}\ \emph {et~al.}(2013)\citenamefont
  {Rahmani}, \citenamefont {Muniz},\ and\ \citenamefont
  {Martin}}]{rahmani-2013}%
  \BibitemOpen
  \bibfield  {author} {\bibinfo {author} {\bibfnamefont {A.}~\bibnamefont
  {Rahmani}}, \bibinfo {author} {\bibfnamefont {R.~A.}\ \bibnamefont {Muniz}},
  \ and\ \bibinfo {author} {\bibfnamefont {I.}~\bibnamefont {Martin}},\ }\href
  {\doibase 10.1103/PhysRevX.3.031008} {\bibfield  {journal} {\bibinfo
  {journal} {Phys. Rev. X}\ }\textbf {\bibinfo {volume} {3}},\ \bibinfo {pages}
  {031008} (\bibinfo {year} {2013})}\BibitemShut {NoStop}%
\bibitem [{\citenamefont {Takatsu}\ \emph {et~al.}(2010)\citenamefont
  {Takatsu}, \citenamefont {Yonezawa}, \citenamefont {Fujimoto},\ and\
  \citenamefont {Maeno}}]{takatsu-2010}%
  \BibitemOpen
  \bibfield  {author} {\bibinfo {author} {\bibfnamefont {H.}~\bibnamefont
  {Takatsu}}, \bibinfo {author} {\bibfnamefont {S.}~\bibnamefont {Yonezawa}},
  \bibinfo {author} {\bibfnamefont {S.}~\bibnamefont {Fujimoto}}, \ and\
  \bibinfo {author} {\bibfnamefont {Y.}~\bibnamefont {Maeno}},\ }\href
  {\doibase 10.1103/PhysRevLett.105.137201} {\bibfield  {journal} {\bibinfo
  {journal} {Phys. Rev. Lett.}\ }\textbf {\bibinfo {volume} {105}},\ \bibinfo
  {pages} {137201} (\bibinfo {year} {2010})}\BibitemShut {NoStop}%
\bibitem [{\citenamefont {Barros}\ \emph {et~al.}(2014)\citenamefont {Barros},
  \citenamefont {Venderbos}, \citenamefont {Chern},\ and\ \citenamefont
  {Batista}}]{kipton-2014}%
  \BibitemOpen
  \bibfield  {author} {\bibinfo {author} {\bibfnamefont {K.}~\bibnamefont
  {Barros}}, \bibinfo {author} {\bibfnamefont {J.~W.~F.}\ \bibnamefont
  {Venderbos}}, \bibinfo {author} {\bibfnamefont {G.-W.}\ \bibnamefont
  {Chern}}, \ and\ \bibinfo {author} {\bibfnamefont {C.~D.}\ \bibnamefont
  {Batista}},\ }\href {\doibase 10.1103/PhysRevB.90.245119} {\bibfield
  {journal} {\bibinfo  {journal} {Phys. Rev. B}\ }\textbf {\bibinfo {volume}
  {90}},\ \bibinfo {pages} {245119} (\bibinfo {year} {2014})}\BibitemShut
  {NoStop}%
\bibitem [{\citenamefont {Chern}\ \emph {et~al.}(2014)\citenamefont {Chern},
  \citenamefont {Rahmani}, \citenamefont {Martin},\ and\ \citenamefont
  {Batista}}]{chern-2014}%
  \BibitemOpen
  \bibfield  {author} {\bibinfo {author} {\bibfnamefont {G.-W.}\ \bibnamefont
  {Chern}}, \bibinfo {author} {\bibfnamefont {A.}~\bibnamefont {Rahmani}},
  \bibinfo {author} {\bibfnamefont {I.}~\bibnamefont {Martin}}, \ and\ \bibinfo
  {author} {\bibfnamefont {C.~D.}\ \bibnamefont {Batista}},\ }\href {\doibase
  10.1103/PhysRevB.90.241102} {\bibfield  {journal} {\bibinfo  {journal} {Phys.
  Rev. B}\ }\textbf {\bibinfo {volume} {90}},\ \bibinfo {pages} {241102}
  (\bibinfo {year} {2014})}\BibitemShut {NoStop}%
\bibitem [{\citenamefont {Venderbos}\ \emph {et~al.}(2012)\citenamefont
  {Venderbos}, \citenamefont {Daghofer}, \citenamefont {van~den Brink},\ and\
  \citenamefont {Kumar}}]{venderbos-2012}%
  \BibitemOpen
  \bibfield  {author} {\bibinfo {author} {\bibfnamefont {J.~W.~F.}\
  \bibnamefont {Venderbos}}, \bibinfo {author} {\bibfnamefont {M.}~\bibnamefont
  {Daghofer}}, \bibinfo {author} {\bibfnamefont {J.}~\bibnamefont {van~den
  Brink}}, \ and\ \bibinfo {author} {\bibfnamefont {S.}~\bibnamefont {Kumar}},\
  }\href {\doibase 10.1103/PhysRevLett.109.166405} {\bibfield  {journal}
  {\bibinfo  {journal} {Phys. Rev. Lett.}\ }\textbf {\bibinfo {volume} {109}},\
  \bibinfo {pages} {166405} (\bibinfo {year} {2012})}\BibitemShut {NoStop}%
\bibitem [{\citenamefont {Chern}(2010)}]{chern-2010}%
  \BibitemOpen
  \bibfield  {author} {\bibinfo {author} {\bibfnamefont {G.-W.}\ \bibnamefont
  {Chern}},\ }\href {\doibase 10.1103/PhysRevLett.105.226403} {\bibfield
  {journal} {\bibinfo  {journal} {Phys. Rev. Lett.}\ }\textbf {\bibinfo
  {volume} {105}},\ \bibinfo {pages} {226403} (\bibinfo {year}
  {2010})}\BibitemShut {NoStop}%
\bibitem [{\citenamefont {Nakatsuji}\ \emph {et~al.}(2006)\citenamefont
  {Nakatsuji}, \citenamefont {Machida}, \citenamefont {Maeno}, \citenamefont
  {Tayama}, \citenamefont {Sakakibara}, \citenamefont {van Duijn},
  \citenamefont {Balicas}, \citenamefont {Millican}, \citenamefont {Macaluso},\
  and\ \citenamefont {Chan}}]{nakatsuji-2006}%
  \BibitemOpen
  \bibfield  {author} {\bibinfo {author} {\bibfnamefont {S.}~\bibnamefont
  {Nakatsuji}}, \bibinfo {author} {\bibfnamefont {Y.}~\bibnamefont {Machida}},
  \bibinfo {author} {\bibfnamefont {Y.}~\bibnamefont {Maeno}}, \bibinfo
  {author} {\bibfnamefont {T.}~\bibnamefont {Tayama}}, \bibinfo {author}
  {\bibfnamefont {T.}~\bibnamefont {Sakakibara}}, \bibinfo {author}
  {\bibfnamefont {J.}~\bibnamefont {van Duijn}}, \bibinfo {author}
  {\bibfnamefont {L.}~\bibnamefont {Balicas}}, \bibinfo {author} {\bibfnamefont
  {J.~N.}\ \bibnamefont {Millican}}, \bibinfo {author} {\bibfnamefont {R.~T.}\
  \bibnamefont {Macaluso}}, \ and\ \bibinfo {author} {\bibfnamefont {J.~Y.}\
  \bibnamefont {Chan}},\ }\href {\doibase 10.1103/PhysRevLett.96.087204}
  {\bibfield  {journal} {\bibinfo  {journal} {Phys. Rev. Lett.}\ }\textbf
  {\bibinfo {volume} {96}},\ \bibinfo {pages} {087204} (\bibinfo {year}
  {2006})}\BibitemShut {NoStop}%
\bibitem [{\citenamefont {Shindou}\ and\ \citenamefont
  {Nagaosa}(2001)}]{shindou-2001}%
  \BibitemOpen
  \bibfield  {author} {\bibinfo {author} {\bibfnamefont {R.}~\bibnamefont
  {Shindou}}\ and\ \bibinfo {author} {\bibfnamefont {N.}~\bibnamefont
  {Nagaosa}},\ }\href {\doibase 10.1103/PhysRevLett.87.116801} {\bibfield
  {journal} {\bibinfo  {journal} {Phys. Rev. Lett.}\ }\textbf {\bibinfo
  {volume} {87}},\ \bibinfo {pages} {116801} (\bibinfo {year}
  {2001})}\BibitemShut {NoStop}%
\bibitem [{\citenamefont {Chen}\ \emph {et~al.}(2014)\citenamefont {Chen},
  \citenamefont {Niu},\ and\ \citenamefont {MacDonald}}]{chen-2014}%
  \BibitemOpen
  \bibfield  {author} {\bibinfo {author} {\bibfnamefont {H.}~\bibnamefont
  {Chen}}, \bibinfo {author} {\bibfnamefont {Q.}~\bibnamefont {Niu}}, \ and\
  \bibinfo {author} {\bibfnamefont {A.~H.}\ \bibnamefont {MacDonald}},\ }\href
  {\doibase 10.1103/PhysRevLett.112.017205} {\bibfield  {journal} {\bibinfo
  {journal} {Phys. Rev. Lett.}\ }\textbf {\bibinfo {volume} {112}},\ \bibinfo
  {pages} {017205} (\bibinfo {year} {2014})}\BibitemShut {NoStop}%
\bibitem [{\citenamefont {Kübler}\ and\ \citenamefont
  {Felser}(2014)}]{kubler-2014}%
  \BibitemOpen
  \bibfield  {author} {\bibinfo {author} {\bibfnamefont {J.}~\bibnamefont
  {Kübler}}\ and\ \bibinfo {author} {\bibfnamefont {C.}~\bibnamefont
  {Felser}},\ }\href {http://stacks.iop.org/0295-5075/108/i=6/a=67001}
  {\bibfield  {journal} {\bibinfo  {journal} {Europhys. Lett.}\ }\textbf
  {\bibinfo {volume} {108}},\ \bibinfo {pages} {67001} (\bibinfo {year}
  {2014})}\BibitemShut {NoStop}%
\bibitem [{\citenamefont {Nayak}\ \emph {et~al.}(2016)\citenamefont {Nayak},
  \citenamefont {Fischer}, \citenamefont {Sun}, \citenamefont {Yan},
  \citenamefont {Karel}, \citenamefont {Komarek}, \citenamefont {Shekhar},
  \citenamefont {Kumar}, \citenamefont {Schnelle}, \citenamefont {K{\"u}bler},
  \citenamefont {Felser},\ and\ \citenamefont {Parkin}}]{nayake-2016}%
  \BibitemOpen
  \bibfield  {author} {\bibinfo {author} {\bibfnamefont {A.~K.}\ \bibnamefont
  {Nayak}}, \bibinfo {author} {\bibfnamefont {J.~E.}\ \bibnamefont {Fischer}},
  \bibinfo {author} {\bibfnamefont {Y.}~\bibnamefont {Sun}}, \bibinfo {author}
  {\bibfnamefont {B.}~\bibnamefont {Yan}}, \bibinfo {author} {\bibfnamefont
  {J.}~\bibnamefont {Karel}}, \bibinfo {author} {\bibfnamefont {A.~C.}\
  \bibnamefont {Komarek}}, \bibinfo {author} {\bibfnamefont {C.}~\bibnamefont
  {Shekhar}}, \bibinfo {author} {\bibfnamefont {N.}~\bibnamefont {Kumar}},
  \bibinfo {author} {\bibfnamefont {W.}~\bibnamefont {Schnelle}}, \bibinfo
  {author} {\bibfnamefont {J.}~\bibnamefont {K{\"u}bler}}, \bibinfo {author}
  {\bibfnamefont {C.}~\bibnamefont {Felser}}, \ and\ \bibinfo {author}
  {\bibfnamefont {S.~S.~P.}\ \bibnamefont {Parkin}},\ }\href
  {http://advances.sciencemag.org/content/2/4/e1501870} {\bibfield  {journal}
  {\bibinfo  {journal} {Sci. Adv.}\ }\textbf {\bibinfo {volume} {2}},\ \bibinfo
  {pages} {e1501870} (\bibinfo {year} {2016})}\BibitemShut {NoStop}%
\bibitem [{\citenamefont {Shahzad}\ and\ \citenamefont
  {Sengupta}(2017)}]{shahzad-2017}%
  \BibitemOpen
  \bibfield  {author} {\bibinfo {author} {\bibfnamefont {M.}~\bibnamefont
  {Shahzad}}\ and\ \bibinfo {author} {\bibfnamefont {P.}~\bibnamefont
  {Sengupta}},\ }\href {\doibase 10.1103/PhysRevB.96.224401} {\bibfield
  {journal} {\bibinfo  {journal} {Phys. Rev. B}\ }\textbf {\bibinfo {volume}
  {96}},\ \bibinfo {pages} {224401} (\bibinfo {year} {2017})}\BibitemShut
  {NoStop}%
\bibitem [{\citenamefont {Pekker}\ \emph {et~al.}(2005)\citenamefont {Pekker},
  \citenamefont {Mukhopadhyay}, \citenamefont {Trivedi},\ and\ \citenamefont
  {Goldbart}}]{pekker-2005}%
  \BibitemOpen
  \bibfield  {author} {\bibinfo {author} {\bibfnamefont {D.}~\bibnamefont
  {Pekker}}, \bibinfo {author} {\bibfnamefont {S.}~\bibnamefont
  {Mukhopadhyay}}, \bibinfo {author} {\bibfnamefont {N.}~\bibnamefont
  {Trivedi}}, \ and\ \bibinfo {author} {\bibfnamefont {P.~M.}\ \bibnamefont
  {Goldbart}},\ }\href {\doibase 10.1103/PhysRevB.72.075118} {\bibfield
  {journal} {\bibinfo  {journal} {Phys. Rev. B}\ }\textbf {\bibinfo {volume}
  {72}},\ \bibinfo {pages} {075118} (\bibinfo {year} {2005})}\BibitemShut
  {NoStop}%
\bibitem [{\citenamefont {Martin}\ and\ \citenamefont
  {Batista}(2008)}]{martin-2008}%
  \BibitemOpen
  \bibfield  {author} {\bibinfo {author} {\bibfnamefont {I.}~\bibnamefont
  {Martin}}\ and\ \bibinfo {author} {\bibfnamefont {C.~D.}\ \bibnamefont
  {Batista}},\ }\href {\doibase 10.1103/PhysRevLett.101.156402} {\bibfield
  {journal} {\bibinfo  {journal} {Phys. Rev. Lett.}\ }\textbf {\bibinfo
  {volume} {101}},\ \bibinfo {pages} {156402} (\bibinfo {year}
  {2008})}\BibitemShut {NoStop}%
\bibitem [{\citenamefont {Motome}\ and\ \citenamefont
  {Furukawa}(1999)}]{motome-1999}%
  \BibitemOpen
  \bibfield  {author} {\bibinfo {author} {\bibfnamefont {Y.}~\bibnamefont
  {Motome}}\ and\ \bibinfo {author} {\bibfnamefont {N.}~\bibnamefont
  {Furukawa}},\ }\href {\doibase 10.1143/JPSJ.68.3853} {\bibfield  {journal}
  {\bibinfo  {journal} {J. Phys. Soc. Jpn.}\ }\textbf {\bibinfo {volume}
  {68}},\ \bibinfo {pages} {3853} (\bibinfo {year} {1999})}\BibitemShut
  {NoStop}%
\bibitem [{\citenamefont {Furukawa}\ and\ \citenamefont
  {Motome}(2004)}]{furukawa-2004}%
  \BibitemOpen
  \bibfield  {author} {\bibinfo {author} {\bibfnamefont {N.}~\bibnamefont
  {Furukawa}}\ and\ \bibinfo {author} {\bibfnamefont {Y.}~\bibnamefont
  {Motome}},\ }\href {\doibase 10.1143/JPSJ.73.1482} {\bibfield  {journal}
  {\bibinfo  {journal} {J. Phys. Soc. Jpn.}\ }\textbf {\bibinfo {volume}
  {73}},\ \bibinfo {pages} {1482} (\bibinfo {year} {2004})}\BibitemShut
  {NoStop}%
\bibitem [{\citenamefont {Ishizuka}\ and\ \citenamefont
  {Motome}(2012)}]{ishizuka-2012}%
  \BibitemOpen
  \bibfield  {author} {\bibinfo {author} {\bibfnamefont {H.}~\bibnamefont
  {Ishizuka}}\ and\ \bibinfo {author} {\bibfnamefont {Y.}~\bibnamefont
  {Motome}},\ }\href {\doibase 10.1103/PhysRevLett.108.257205} {\bibfield
  {journal} {\bibinfo  {journal} {Phys. Rev. Lett.}\ }\textbf {\bibinfo
  {volume} {108}},\ \bibinfo {pages} {257205} (\bibinfo {year}
  {2012})}\BibitemShut {NoStop}%
\bibitem [{\citenamefont {\ifmmode~\mbox{\c{S}}\else \c{S}\fi{}en}\ \emph
  {et~al.}(2006)\citenamefont {\ifmmode~\mbox{\c{S}}\else \c{S}\fi{}en},
  \citenamefont {Alvarez}, \citenamefont {Motome}, \citenamefont {Furukawa},
  \citenamefont {Sergienko}, \citenamefont {Schulthess}, \citenamefont
  {Moreo},\ and\ \citenamefont {Dagotto}}]{alvarez-2006}%
  \BibitemOpen
  \bibfield  {author} {\bibinfo {author} {\bibfnamefont {C.}~\bibnamefont
  {\ifmmode~\mbox{\c{S}}\else \c{S}\fi{}en}}, \bibinfo {author} {\bibfnamefont
  {G.}~\bibnamefont {Alvarez}}, \bibinfo {author} {\bibfnamefont
  {Y.}~\bibnamefont {Motome}}, \bibinfo {author} {\bibfnamefont
  {N.}~\bibnamefont {Furukawa}}, \bibinfo {author} {\bibfnamefont {I.~A.}\
  \bibnamefont {Sergienko}}, \bibinfo {author} {\bibfnamefont {T.~C.}\
  \bibnamefont {Schulthess}}, \bibinfo {author} {\bibfnamefont
  {A.}~\bibnamefont {Moreo}}, \ and\ \bibinfo {author} {\bibfnamefont
  {E.}~\bibnamefont {Dagotto}},\ }\href {\doibase 10.1103/PhysRevB.73.224430}
  {\bibfield  {journal} {\bibinfo  {journal} {Phys. Rev. B}\ }\textbf {\bibinfo
  {volume} {73}},\ \bibinfo {pages} {224430} (\bibinfo {year}
  {2006})}\BibitemShut {NoStop}%
\bibitem [{\citenamefont {Ishizuka}\ and\ \citenamefont
  {Motome}(2013)}]{ishizuka-2013}%
  \BibitemOpen
  \bibfield  {author} {\bibinfo {author} {\bibfnamefont {H.}~\bibnamefont
  {Ishizuka}}\ and\ \bibinfo {author} {\bibfnamefont {Y.}~\bibnamefont
  {Motome}},\ }\href {\doibase 10.1103/PhysRevB.88.081105} {\bibfield
  {journal} {\bibinfo  {journal} {Phys. Rev. B}\ }\textbf {\bibinfo {volume}
  {88}},\ \bibinfo {pages} {081105} (\bibinfo {year} {2013})}\BibitemShut
  {NoStop}%
\bibitem [{\citenamefont {Ishizuka}\ and\ \citenamefont
  {Motome}(2015{\natexlab{a}})}]{ishizuka-2015}%
  \BibitemOpen
  \bibfield  {author} {\bibinfo {author} {\bibfnamefont {H.}~\bibnamefont
  {Ishizuka}}\ and\ \bibinfo {author} {\bibfnamefont {Y.}~\bibnamefont
  {Motome}},\ }\href {\doibase 10.1103/PhysRevB.91.085110} {\bibfield
  {journal} {\bibinfo  {journal} {Phys. Rev. B}\ }\textbf {\bibinfo {volume}
  {91}},\ \bibinfo {pages} {085110} (\bibinfo {year}
  {2015}{\natexlab{a}})}\BibitemShut {NoStop}%
\bibitem [{\citenamefont {{Kumar, S.}}\ and\ \citenamefont {{Majumdar,
  P.}}(2006)}]{kumar-2006}%
  \BibitemOpen
  \bibfield  {author} {\bibinfo {author} {\bibnamefont {{Kumar, S.}}}\ and\
  \bibinfo {author} {\bibnamefont {{Majumdar, P.}}},\ }\href {\doibase
  10.1140/epjb/e2006-00173-2} {\bibfield  {journal} {\bibinfo  {journal} {Eur.
  Phys. J. B}\ }\textbf {\bibinfo {volume} {50}},\ \bibinfo {pages} {571}
  (\bibinfo {year} {2006})}\BibitemShut {NoStop}%
\bibitem [{\citenamefont {Mukherjee}\ \emph {et~al.}(2015)\citenamefont
  {Mukherjee}, \citenamefont {Patel}, \citenamefont {Bishop},\ and\
  \citenamefont {Dagotto}}]{anamitra-2015}%
  \BibitemOpen
  \bibfield  {author} {\bibinfo {author} {\bibfnamefont {A.}~\bibnamefont
  {Mukherjee}}, \bibinfo {author} {\bibfnamefont {N.~D.}\ \bibnamefont
  {Patel}}, \bibinfo {author} {\bibfnamefont {C.}~\bibnamefont {Bishop}}, \
  and\ \bibinfo {author} {\bibfnamefont {E.}~\bibnamefont {Dagotto}},\ }\href
  {\doibase 10.1103/PhysRevE.91.063303} {\bibfield  {journal} {\bibinfo
  {journal} {Phys. Rev. E}\ }\textbf {\bibinfo {volume} {91}},\ \bibinfo
  {pages} {063303} (\bibinfo {year} {2015})}\BibitemShut {NoStop}%
\bibitem [{\citenamefont {Kumar}\ and\ \citenamefont
  {Majumdar}(2006)}]{majumdar-2006}%
  \BibitemOpen
  \bibfield  {author} {\bibinfo {author} {\bibfnamefont {S.}~\bibnamefont
  {Kumar}}\ and\ \bibinfo {author} {\bibfnamefont {P.}~\bibnamefont
  {Majumdar}},\ }\href {\doibase 10.1103/PhysRevLett.96.016602} {\bibfield
  {journal} {\bibinfo  {journal} {Phys. Rev. Lett.}\ }\textbf {\bibinfo
  {volume} {96}},\ \bibinfo {pages} {016602} (\bibinfo {year}
  {2006})}\BibitemShut {NoStop}%
\bibitem [{\citenamefont {Kumar}\ and\ \citenamefont
  {Majumdar}(2005)}]{kumar-2005}%
  \BibitemOpen
  \bibfield  {author} {\bibinfo {author} {\bibfnamefont {S.}~\bibnamefont
  {Kumar}}\ and\ \bibinfo {author} {\bibfnamefont {P.}~\bibnamefont
  {Majumdar}},\ }\href {\doibase 10.1103/PhysRevLett.94.136601} {\bibfield
  {journal} {\bibinfo  {journal} {Phys. Rev. Lett.}\ }\textbf {\bibinfo
  {volume} {94}},\ \bibinfo {pages} {136601} (\bibinfo {year}
  {2005})}\BibitemShut {NoStop}%
\bibitem [{\citenamefont {Yamanaka}\ \emph {et~al.}(1998)\citenamefont
  {Yamanaka}, \citenamefont {Koshibae},\ and\ \citenamefont
  {Maekawa}}]{yamanaka-1998}%
  \BibitemOpen
  \bibfield  {author} {\bibinfo {author} {\bibfnamefont {M.}~\bibnamefont
  {Yamanaka}}, \bibinfo {author} {\bibfnamefont {W.}~\bibnamefont {Koshibae}},
  \ and\ \bibinfo {author} {\bibfnamefont {S.}~\bibnamefont {Maekawa}},\ }\href
  {\doibase 10.1103/PhysRevLett.81.5604} {\bibfield  {journal} {\bibinfo
  {journal} {Phys. Rev. Lett.}\ }\textbf {\bibinfo {volume} {81}},\ \bibinfo
  {pages} {5604} (\bibinfo {year} {1998})}\BibitemShut {NoStop}%
\bibitem [{\citenamefont {Agterberg}\ and\ \citenamefont
  {Yunoki}(2000)}]{agterberg-2000}%
  \BibitemOpen
  \bibfield  {author} {\bibinfo {author} {\bibfnamefont {D.~F.}\ \bibnamefont
  {Agterberg}}\ and\ \bibinfo {author} {\bibfnamefont {S.}~\bibnamefont
  {Yunoki}},\ }\href {\doibase 10.1103/PhysRevB.62.13816} {\bibfield  {journal}
  {\bibinfo  {journal} {Phys. Rev. B}\ }\textbf {\bibinfo {volume} {62}},\
  \bibinfo {pages} {13816} (\bibinfo {year} {2000})}\BibitemShut {NoStop}%
\bibitem [{\citenamefont {Ishizuka}\ and\ \citenamefont
  {Motome}(2015{\natexlab{b}})}]{hiroaki-2015}%
  \BibitemOpen
  \bibfield  {author} {\bibinfo {author} {\bibfnamefont {H.}~\bibnamefont
  {Ishizuka}}\ and\ \bibinfo {author} {\bibfnamefont {Y.}~\bibnamefont
  {Motome}},\ }\href {\doibase 10.1103/PhysRevB.92.024415} {\bibfield
  {journal} {\bibinfo  {journal} {Phys. Rev. B}\ }\textbf {\bibinfo {volume}
  {92}},\ \bibinfo {pages} {024415} (\bibinfo {year}
  {2015}{\natexlab{b}})}\BibitemShut {NoStop}%
\end{thebibliography}%

    \end{document}